\documentclass[journal]{IEEEtran}
\usepackage{amsfonts}
\usepackage{amsmath}
\usepackage{amssymb}
\usepackage{lscape}
\usepackage{epsf}
\usepackage{graphicx}

\newtheorem{theorem}{\indent Theorem}[section]

\newtheorem{EXAMPLE}{\indent Example}[section]
\newtheorem{definition}{\indent Definition}[section]

\newcommand{\code}{{\mathcal{C}}}

\newcommand{\cL}{{\mathcal{L}}}
\newcommand{\cV}{{\mathcal{V}}}
\newcommand{\cG}{{\mathcal{G}}}
\newcommand{\cH}{{\mathbf{H}}}

\newcommand{\cA}{{\mathcal{A}}}
\newcommand{\cB}{{\mathcal{B}}}
\newcommand{\cD}{{\mathcal{D}}}
\newcommand{\cF}{{\mathcal{F}}}
\newcommand{\cI}{{\mathcal{I}}}
\newcommand{\cJ}{{\mathcal{J}}}
\newcommand{\cM}{{\mathcal{M}}}

\newcommand{\cK}{{\mathcal{K}}}
\newcommand{\cP}{{\mathcal{P}}}
\newcommand{\cQ}{{\mathcal{Q}}}
\newcommand{\cR}{{\mathcal{R}}}
\newcommand{\cS}{{\mathcal{S}}}
\newcommand{\cT}{{\mathcal{T}}}
\newcommand{\cU}{{\mathcal{U}}}
\newcommand{\cX}{{\mathcal{X}}}
\newcommand{\cY}{{\mathcal{Y}}}
\newcommand{\cYY}{{\mathcal{Y}}}

\newcommand{\ip}{{\mathrm{ip}}}
\newcommand{\op}{{\mathrm{op}}}

\newcommand{\bldb}{{\mbox{\boldmath $b$}}}
\newcommand{\bldbb}{{\mbox{\scriptsize \boldmath $b$}}}
\newcommand{\bldc}{{\mbox{\boldmath $c$}}}

\newcommand{\bldd}{{\mbox{\boldmath $d$}}}
\newcommand{\blddd}{{\mbox{\scriptsize \boldmath $d$}}}

\newcommand{\bldf}{{\mbox{\boldmath $f$}}}

\newcommand{\bldg}{{\mbox{\boldmath $g$}}}
\newcommand{\bldgg}{{\mbox{\scriptsize \boldmath $g$}}}

\newcommand{\bldkappa}{{\mbox{\boldmath $\kappa$}}}
\newcommand{\bldsubkappa}{{\mbox{\scriptsize \boldmath $\kappa$}}}
\newcommand{\bldm}{{\mbox{\boldmath $m$}}}
\newcommand{\bldp}{{\mbox{\boldmath $p$}}}
\newcommand{\bldq}{{\mbox{\boldmath $q$}}}

\newcommand{\blds}{{\mbox{\boldmath $s$}}}

\newcommand{\bldt}{{\mbox{\boldmath $t$}}}

\newcommand{\bldP}{{\mbox{\boldmath $P$}}}

\newcommand{\bldv}{{\mbox{\boldmath $v$}}}
\newcommand{\bldV}{{\mbox{\boldmath $V$}}}

\newcommand{\bldw}{{\mbox{\boldmath $w$}}}

\newcommand{\bldww}{{\mbox{\scriptsize \boldmath $w$}}}
\newcommand{\bldx}{{\mbox{\boldmath $x$}}}
\newcommand{\bldxx}{{\mbox{\scriptsize \boldmath $x$}}}
\newcommand{\bldxi}{{\mbox{\boldmath $\xi$}}}
\newcommand{\bldXi}{{\mbox{\boldmath $\Xi$}}}

\newcommand{\bldy}{{\mbox{\boldmath $y$}}}

\newcommand{\bldz}{{\mbox{\boldmath $z$}}}
\newcommand{\bldlambda}{{\mbox{\boldmath $\lambda$}}}

\newcommand{\zeros}{{\mbox{\boldmath $0$}}}
\newcommand{\rrr}{\mathfrak{R}}%

\newcommand{\bldzeros}{{\mbox{\boldmath $0$}}}

\newcommand{\bldeta}{{\mbox{\boldmath $\eta$}}}

    \def\squarebox#1{\hbox to #1{\hfill\vbox to #1{\vfill}}}

\newlength{\Algwidth}

\begin{document}

\title{A Unified Framework for Linear-Programming Based Communication Receivers 
\thanks{%
     This work was supported in part by Science Foundation Ireland (Grant No. 07/SK/I1252b), in part by the Institute of Advanced Studies, University of Bologna (ISA-ESRF Fellowship), and in part by the Swiss National Science Foundation (Grant No. 113251). The material in this paper was presented in part at the 46-th International Allerton Conference on Communication, Control and Computing, Monticello, Illinois, September 2008.  
    \newline 
    M. F. Flanagan was with DEIS, University of Bologna, Via Venezia 52, 47023 Cesena, Italy, and with Institut f\"ur Mathematik, University of Z\"urich, Winterthurerstrasse 190, CH-8057 Z\"urich, Switzerland. He is now with the School of Electrical, Electronic and Communications Engineering, University College Dublin, Belfield, Dublin 4, Ireland (email:mark.flanagan@ieee.org).
    }        
}
\author{Mark F. Flanagan,~\IEEEmembership{Senior Member,~IEEE}} 

\maketitle

\vspace{-0.5cm}
\begin{abstract}
It is shown that a large class of communication systems which admit a sum-product algorithm (SPA) based receiver also admit a corresponding linear-programming (LP) based receiver. The two receivers have a relationship defined by the local structure of the underlying graphical model, and are inhibited by the same phenomenon, which we call \emph{pseudoconfigurations}. This concept is a generalization of the concept of \emph{pseudocodewords} for linear codes. It is proved that the LP receiver has the `maximum likelihood certificate' property, and that the receiver output is the lowest cost pseudoconfiguration. Equivalence of graph-cover pseudoconfigurations and linear-programming pseudoconfigurations is also proved. A concept of \emph{system pseudodistance} is defined which generalizes the existing concept of pseudodistance for binary and nonbinary linear codes. It is demonstrated how the LP design technique may be applied to the problem of joint equalization and decoding of coded transmissions over a frequency selective channel, and a simulation-based analysis of the error events of the resulting LP receiver is also provided. For this particular application, the proposed LP receiver is shown to be competitive with other receivers, and to be capable of outperforming turbo equalization in bit and frame error rate performance.
\end{abstract}


\newcounter{mytempeqncnt}

\section{Introduction}

The decoding algorithms for some of the best known classes of error-correcting code to date, namely concatenated (``turbo'') codes \cite{Berrou} and low-density parity check (LDPC) codes \cite{Gallager}, have been shown to be instances of a much more general algorithm called the \emph{sum-product algorithm} (SPA) \cite{Wiberg,Aji, Kschischang}. This algorithm solves the general problem of marginalizing a product of functions which take values in a semiring $\cR$. In the communications context, $\cR$ is equal to $\mathbb{R}_{\geq 0}$, the set of nonnegative real numbers, and the maximization of each marginal function minimizes the error rate for each symbol, under the assumption that the system factor graph is a tree \cite{Kschischang}. It has been recognized that many diverse situations may allow the use of SPA based reception \cite{Unified_design}, including joint iterative equalization and decoding (or \emph{turbo equalization}) \cite{turbo_eq}, joint iterative equalization and multiuser detection (MUD) \cite{Li_Poor},  and joint source-channel decoding \cite{Goertz}. 

Recently, a linear-programming (LP) based approach to decoding linear (and especially LDPC) codes was developed for binary \cite{Feldman-thesis, Feldman} and nonbinary coding frameworks \cite{FSBG_SCC, FSBG_journal}. The concept of \emph{pseudocodeword} proved important in the performance analysis of both LP and SPA based decoders \cite{FKKR, KV-characterization, KV-IEEE-IT}. Also, linear-programming decoders for irregular repeat-accumulate (IRA) codes and turbo codes were described in \cite{Feldman_turbo_IRA}. While the complexity of LP decoding is conjectured to be higher than for SPA decoding, the LP decoder has many analytical advantages, such as the property that a codeword output by the LP is always the maximum likelihood (ML) codeword, and the equivalence of different pseudocodeword concepts in the LP and SPA domains \cite{Feldman, FSBG_journal}. For the case of LDPC codes, tight connections were observed between the LP decoding and min-sum decoding frameworks \cite{Vontobel_min_sum_LP}.

Recently, some authors have considered use of similar linear-programming techniques in applications beyond coding. An LP-based method for low-complexity joint equalization and decoding of LDPC coded transmissions over the magnetic recording channel was proposed in \cite{Siegel}. In this work, the problem of ML joint detection, which may be expressed as an integer quadratic program, is converted into a linear programming relaxation of a binary-constrained problem. In the case where there is no coding, it was shown in \cite{Siegel} that for a class of channels designated as \emph{proper} channels, the LP solution matches the ML solution at all values of signal to noise ratio (SNR); however, for some channels which are not proper, the system evinces a frame error rate \emph{floor} effect. The work of \cite{Cohen} considered an LP decoder which incorporates nonuniform priors into the original decoding polytope of \cite{Feldman}, and also application of an LP decoder to transmissions over a channel with memory, namely the non-ergodic Polya channel. In both \cite{Siegel} and \cite{Cohen}, performance analysis proved difficult for the case where the channel has memory.

In this paper it is shown that the problem of maximizing a product of $\cR$-valued functions is amenable to an approximate (suboptimal) solution using an LP relaxation, under two conditions: first, that the semiring $\cR$ corresponds to $\mathbb{R}_{\geq 0}$ under real addition and multiplication, and second, that all factor nodes of degree greater than one are indicator functions for a local behavior. Fortunately, these conditions are satisfied by a large number of practical communication receiver design problems. Interestingly, the LP exhibits a ``separation effect'' in the sense that degree-$1$ factor nodes in the factor graph contribute the cost function, and the remaining nodes determine the LP constraint set. This distinction is somewhat analogous to the case of SPA-based reception where degree-$1$ factor nodes contribute initial messages exactly once, and all other nodes update their messages periodically. Our LP receiver generalizes the LP \emph{decoders} of \cite{Feldman, FSBG_journal,Feldman_turbo_IRA}. A general design methodology emerges, parallel to that of SPA receiver design:
\begin{enumerate} 
\item 
Write down the global function for the transmitter-channel combination. This is a function proportional to the probability mass function of the transmitter configuration conditioned on the received observations.
\item
Draw the factor graph corresponding to the global function.
\item
Read the LP variables and constraints directly from the factor graph.
\end{enumerate} 

The proposed framework applies to any system with a finite number of transmitter configurations, and also allows for treatment of ``hidden'' (latent) state variables (as was done for the SPA receiver case in \cite{Wiberg}). It allows for a systematic treatment of variables with known values (e.g. known initial/final channel states or pilot symbols). Incorporation of priors for any subset of transmitter variables is straightforward, and thus the polytopes of \cite[Section II]{Cohen} follow as simple special cases of our framework. Our framework allows derivation of LP decoders also for tail-biting trellis (TBT) codes; in this case the relevant pseudoconfigurations correspond to the TBT pseudocodewords as defined in \cite{FKKR}. It is proved that the LP receiver error events, which we characterize as a set of \emph{linear-programming pseudoconfigurations}, are equivalent to the set of \emph{graph-cover pseudoconfigurations}, which are linked to error events in the corresponding SPA receiver. Furthermore, we define a general concept of \emph{pseudodistance} for the transmission system, which generalizes the existing concept of pseudodistance for binary and nonbinary linear codes to the case of the general LP receiver.   

In order to illustrate the LP receiver design methodology outlined above, we provide a step-by-step derivation of an LP receiver which performs joint equalization and decoding of coded transmissions over a frequency selective channel. We then provide a simulation study of a simple case of this receiver, including error rate results and pseudodistance spectra, together with a complete description of error events at low values of pseudodistance. Performance results are also presented for a low-density code transmitted over an intersymbol interference channel. We note that a similar line of work is considered in \cite{Kim_Pfister, Kim_Pfister2}; the LP presented therein is equivalent to the one we derive in Section \ref{sec:equalization_decoding}, except that it does not deal with the case of known states in the trellis. The proposed LP receiver is capable of handling channels which are problematic for competitive LP-based techniques, and is shown to have error rate performance outperforming that of turbo equalization over some channels.

This paper is organized as follows. Section \ref{Prob_statement} introduces the general problem to be solved, along with appropriate notations, and Section \ref{max_prod_fn_LP} develops a general linear program which solves this problem. Section \ref{sec:efficient_LP_theoretical} introduces an efficient linear program which provides a suboptimal solution, and Section \ref{sec:efficient_LP} develops an equivalent program with a lower description complexity. Section \ref{sec:PCFs} introduces general concepts of system pseudoconfigurations and pseudodistance. Section \ref{sec:equalization_decoding} provides a detailed development of an LP receiver which performs joint equalization and decoding, and Section \ref{sec:Hamming_Proakis} presents a detailed simulation-based analysis of this receiver for the case of binary-coded transmissions over an intersymbol interference channel. 

\section{Problem Statement and Notations}
\label{Prob_statement}
We begin by introducing some definitions and notation. Suppose that we have variables $x_i$, $i \in \cI$, where $\cI$ is a finite set, and the variable $x_i$ lies in the finite set $\cA_i$ for each $i \in \cI$. Let $\bldx = ( x_i )_{i\in\cI}$\footnote{All vectors in the paper are \emph{row} vectors; also, the notation $(v_t)_{t \in \cT}$ denotes a vector whose entries are equal to $\{v_t: \, t \in\cT\}$ with respect to some fixed ordering on the elements of $\cT$.}; then $\bldx$ is called a \emph{configuration}, and the Cartesian product $\cA = \prod_{i \in\cI} \cA_i$ is called the \emph{configuration space}. Suppose now that we wish to find the configuration $\bldx \in \cA$ which maximizes the product of real-valued functions 
\begin{equation}
u\left(\bldx\right)=\prod_{j\in \cJ}f_{j}\left(\bldx_{j}\right)\label{eq:factorization_of_global_function}
\end{equation}
where $\cJ$ is a finite set, $\bldx_j = ( x_i )_{i\in\cI_j}$ and $\cI_j \subseteq \cI$ for each $j \in \cJ$. We define the \emph{optimum} configuration $\bldx_\mathrm{opt}$ to be the configuration $\bldx \in \cA$ which maximizes (\ref{eq:factorization_of_global_function}). The function $u(\bldx)$ is called the \emph{global function} \cite{Kschischang}. In the communication receiver design context, the global function is taken to be any monotonically increasing function of the probability mass function of some set of transmitter-channel variables (information bits, coded symbols, state variables etc.) conditioned on the received observations. Maximization of the global function therefore corresponds to \emph{maximum a posteriori} (MAP) reception\footnote{Note that in most cases of practical importance, a single $\bldx_\mathrm{opt}$ maximizes (\ref{eq:factorization_of_global_function}) with probability $1$; henceforth, we will assume a unique $\bldx_\mathrm{opt}$.}. As we shall see, the key to solving this optimization problem via a low-complexity LP is that the factors $f_j$ in the factorization (\ref{eq:factorization_of_global_function}) each have a \emph{small} number of arguments, i.e., $|\cI_j|$ is small for each $j \in \cJ$.

The factor graph for the global function $u(\bldx)$ and its factorization (\ref{eq:factorization_of_global_function}) is a (bipartite) graph defined as follows. There is a variable node for each variable $x_{i}$ ($i \in \cI$) and a factor node for each factor $f_{j}$ ($j \in \cJ$). An edge connects variable node $x_{i}$ to factor node $f_{j}$ if and only if $x_{i}$ is an argument of $f_{j}$. Note that for any $j\in\cJ$, $\cI_j$ is the set of $i\in\cI$ for which $x_{i}$ is an argument of $f_{j}$. Also, for any $i\in\cI$, the set of $j\in\cJ$ for which $x_{i}$ is an argument of $f_{j}$ is denoted $\cJ_i$. 

Let $\cL \subseteq \cJ$ denote the set of all $j \in \cJ$ such that factor node $f_j$ is an indicator function for some local behavior $\cB_j$, i.e.,
\begin{equation}
f_{j}\left(\bldx_{j}\right) = \mathbb{I}(\bldx_{j} \in \cB_j) \quad \forall j\in \cL 
\label{eq:indicators_for_factor_nodes}
\end{equation}
where the indicator function for the logical predicate $a$ is defined by 
\[
\mathbb{I}(a) = \left\{ \begin{array}{cc}
1 & \textrm{ if } a \textrm{ is true } \\
0 & \textrm{ otherwise. }\end{array}\right.
\]
In the communication receiver design application, the set $\cL$ comprises constraints such as parity-check constraints and state-space constraints, and also may account for variables with known values (pilot symbols, known states). 

Note that we write any $\bldv\in\cB_j$ as $\bldv = ( v_i )_{i\in\cI_j}$, i.e., $\bldv$ is indexed by $\cI_j$. Also we define the \emph{global behavior} $\cB$ as follows: for any $\bldx \in\cA$, we have $\bldx \in\cB$ if and only if $\bldx_j \in\cB_j$ for every $j\in\cL$. The configuration $\bldx \in \cA$ is said to be \emph{valid} if and only if $\bldx \in \cB$.

Next define $\cY$ to be the set of indices of variable nodes which have neighbours \emph{not} belonging to $\cL$, i.e.,
\[
\cY = \{ i \in \cI : \, \exists j \in \cJ_i \backslash \cL \} \; .
\]
We assume that for every $j\in J_i \backslash \cL$, the factor node $f_j$ has degree equal to one. This allows us to define, for each $i\in\cY$,
\[
h_{i}\left(x_{i}\right) = \prod_{j\in J_i \backslash \cL } f_j\left(x_{i}\right) \, .
\]
In the communication receiver design context, the set $\cY$ corresponds to the set of \emph{observables}, i.e., the set of variables for which noisy observations are available, and each $f_j(x_i)$ represents the probability (density) of the symbol $x_i$ conditioned on the corresponding received observation(s).

So, without loss of generality we may write 
\begin{equation}
u\left(\bldx\right)=\prod_{i\in \cY}h_{i}\left(x_{i}\right) \cdot \prod_{j\in \cL}f_{j}\left(\bldx_{j}\right) \; .
\label{eq:factorization_of_global_function2}
\end{equation}

We assume that the function $h_{i}\left(x_{i}\right)$ is positive-valued for each $i\in\cY$. Also, denoting the Cartesian product $\cA_{\cYY} = \prod_{i \in\cY} \cA_i$, we define the projection  
\[
\bldP_{\cYY} : \, \cA \longrightarrow \cA_{\cYY} \quad \mbox{such that} \quad \bldP_{\cYY}\left(\bldx\right) = ( x_i )_{i\in\cY} \, .
\]
This function simply maps any configuration into the configuration subset consisting only of the observables. Also, we adopt the notation $\bldx_{\cYY} = ( x_i )_{i\in\cY}$ for elements of $\cA_{\cYY}$ (i.e., vectors of observables). 

We assume that the mapping $\bldP_{\cYY}$ is injective on $\cB$, i.e., if $\bldx_1, \bldx_2 \in \cB$ and $\bldP_{\cYY}(\bldx_1) = \bldP_{\cYY}(\bldx_2)$, then $\bldx_1 = \bldx_2$. This corresponds to a `well-posed' problem. Note that in the communication receiver design context, observations (or ``channel information'') may only be contributed through the nodes $x_i$ for $i \in \cY$. Therefore, failure of the injectivity property in the communications context would mean that one particular set of channel inputs could correspond to two different transmit information sets, which would reflect badly on system design.

\section{Maximization of the Global Function by Linear Programming}
\label{max_prod_fn_LP}
Using \eqref{eq:indicators_for_factor_nodes} and \eqref{eq:factorization_of_global_function2}, we may write 
\begin{eqnarray*}
\bldx_\mathrm{opt} & = & \arg \max_{\bldxx \in \cA} \left( \prod_{i\in \cY}h_{i}\left(x_{i}\right) \cdot \prod_{j\in \cL}f_{j}\left(\bldx_{j}\right) \right) \\
 & = & \arg \max_{\bldxx \in \cB} \sum_{i\in \cY} \log h_{i}\left(x_{i}\right) \; .
\end{eqnarray*}
For each $i \in \cI$, $\alpha \in \cA_i$, let $\bldxi_i (\alpha) = ( \mathbb{I}(\gamma = \alpha) )_{\gamma \in \cA_i }$, i.e., $\bldxi_i(\alpha)$ is a real vector of length $|\cA_i|$ which acts as an `indicator vector' for the value $\alpha \in \cA_i$. Building on this, for $\bldx_{\cYY} \in \cA_{\cYY}$ we define the indicator vector $\bldXi(\bldx_{\cYY}) = ( \bldxi_i(x_i) )_{i \in\cY}$, which is the concatenation of the individual indicator vectors for each of the elements of $\bldx_{\cYY}$. It is easy to see that $\bldXi$ is an injective function on $\cA_{\cYY}$.

Next, we define the vector $\boldsymbol{\lambda}$ according to
\[
\boldsymbol{\lambda} = ( \boldsymbol{\lambda}_i )_{i \in \cYY} \quad \mbox{where} \quad \boldsymbol{\lambda}_i = ( \lambda_i^{(\alpha)} )_{\alpha \in \cA_i} \quad \forall i \in \cY \; ,
\]
where $\lambda_i^{(\alpha)} = \log h_i(\alpha)$ for each $i \in \cYY$, $\alpha \in \cA_i$. This allows us to develop the formulation of the optimum configuration as
\begin{eqnarray}
\bldx_\mathrm{opt} & = & \arg \max_{\bldxx \in \cB} \sum_{i\in \cY} \log h_{i}\left(x_{i}\right) \nonumber \\
 & = & \arg \max_{\bldxx \in \cB} \sum_{i\in \cY} \boldsymbol{\lambda}_i \bldxi_i (x_i) ^T \nonumber \\
 & = & \arg \max_{\bldxx \in \cB} \boldsymbol{\lambda} \bldXi (\bldP_{\cYY} (\bldx)) ^T \; ,
\label{eq:opt_problem_LP1}
\end{eqnarray}
where in the second line we have used the fact that the inner product ``sifts'' the value $\lambda_i^{(x_i)} = \log h_i(x_i)$ out of the vector $\boldsymbol{\lambda}_i$, and the third line we have expressed the sum of inner products as a single inner product of the corresponding pair of concatenated vectors. Note that the optimization has reduced to the maximization of an inner product of vectors, where the first vector derives only from observations (or ``channel information'') and the second vector derives only from the global behavior (the set of valid configurations). For any vector $\bldg$ of the same dimension as $\boldsymbol{\lambda}$, we adopt the notation
\[
\bldg = ( \bldg_i )_{i \in \cYY} \quad \mbox{where} \quad \bldg_i = ( g_i^{(\alpha)} )_{\alpha \in \cA_i} \quad \forall i \in \cY \; .
\]
Then the maximization problem (\ref{eq:opt_problem_LP1}) may then be recast as a linear program {\bf LP1} as shown below.

\vspace{1pc}
\hspace{-3mm}\fbox{
\begin{minipage}{0.465\textwidth}
{\bf LP1: Optimum Configuration}

\vspace{1mm}

{\bf Cost Function:} $\boldsymbol{\lambda} \bldg ^T$

{\bf Constraints (Polytope $\cK_{\cYY}(\cB)$):} The cost function is maximized over the convex hull of all points corresponding to valid configurations:
\begin{equation}
\bldg \in \cK_{\cYY}(\cB) = \mathrm{conv} \big\{ \bldXi \left( \bldP_{\cYY}\left(\bldx\right) \right) : \, \bldx\in\cB \big\} \; .
\label{eq:convex_hull_LP1}
\end{equation}
{\bf Receiver Output:}
\begin{equation}
\bldx_\mathrm{opt} = \bldP_{\cYY}^{-1} \left( \bldXi^{-1} (\bldg_\mathrm{opt}) \right)
\label{eq:xopt_LP1}
\end{equation}
\end{minipage}}
\vspace{1pc}

The ``polytope of valid configurations'' $\cK_{\cYY}(\cB)$ generalizes the ``codeword polytope'' defined in \cite{Feldman} and \cite{FSBG_journal} in the context of binary and nonbinary linear codes, respectively. The linear program {\bf LP1} has constraint complexity exponential in the number of LP variables, rendering it unsuitable for practical application.

\section{LP Relaxation}
\label{sec:efficient_LP_theoretical}

In order to reduce the description complexity of {\bf LP1}, we introduce auxiliary variables whose constraints, along with those of the elements of the vector $\bldg$ defined previously, will form the relaxed LP problem. We introduce auxiliary variables $p_{j,\bldbb}$ for each $j\in\cL$, $\bldb \in \cB_j$, and we form the vector
\[
\bldp = \left( \bldp_j \right)_{j \in \cL} \; \mbox{ where } \; \bldp_j = \left( p_{j,\bldbb} \right)_{\bldbb \in \cB_j} \: \forall j \in \cL \; .
\]
Also, we define the vector $\bar{\bldg}$ as an extension of $\bldg$ via $\bar{\bldg} = ( \bldg_i )_{i \in \cI}$ where $\bldg_i = ( g_i^{(\alpha)} )_{\alpha \in \cA_i}$ for each $i \in \cI$ (recall that $\bldg = ( \bldg_i )_{i \in \cY}$). For $\bldx \in \cA$ we define the indicator vector $\bar{\bldXi}(\bldx) = ( \bldxi_i(x_i) )_{i \in\cI}$; the function $\bar{\bldXi}$ is injective on $\cA$.

The new LP optimizes the cost function $\boldsymbol{\lambda} \bldg ^T$ over the polytope $\cQ$ defined with respect to variables $\bar{\bldg}$ and $\bldp$, as shown in the following.

\vspace{1pc}
\hspace{-3mm}\fbox{
\begin{minipage}{0.465\textwidth}
{\bf LP2: Efficient Relaxation}

\vspace{1mm}

{\bf Cost Function:} $\boldsymbol{\lambda} \bldg ^T = \mathbb{E}_{\bldgg}\, \log \prod_{i \in \cY} h_i(x_i)$

{\bf Constraints (Polytope $\cQ$):} 
\begin{equation}
\forall j \in \cL, \; \forall \bldb \in \cB_j,  \quad  p_{j,\bldbb} \ge 0 \; ,
\label{eq:equation-polytope-4} 
\end{equation} 
\begin{equation}
\forall j \in \cL, \quad \sum_{\bldbb \in \cB_j} p_{j,\bldbb} = 1 \; ,
\label{eq:equation-polytope-5} 
\end{equation}  
\begin{equation}
\forall j \in \cL, \; \forall i \in \cI_j, \; \forall \alpha \in \cA_i, \quad g_i^{(\alpha)} = \sum_{\bldbb \in \cB_j, \; b_i=\alpha} p_{j,\bldbb} \; .
\label{eq:equation-polytope-theoretical} 
\end{equation} 
{\bf Receiver Output:}
\begin{equation}
\left\{ \begin{array}{cc}
\bldx_\mathrm{out} = \bldP_{\cYY}^{-1} \left( \bar{\bldXi}^{-1} (\bar{\bldg}_\mathrm{out}) \right) & \textrm{ if } (\bar{\bldg}_\mathrm{out}, \bldp) \textrm{ is integral } \\
{\tt FAILURE} & \textrm{ otherwise. }\end{array}\right.
\label{eq:LP2_output} 
\end{equation} 
\end{minipage}}
\vspace{1pc}

{\bf LP2} is a direct generalization of the LP of \cite{Feldman} to the case of arbitrary behavioral contraints. It comprises $\sum_{j \in \cL} |\cB_j| + \sum_{i \in \cI} |\cA_i|$ variables and $|\cL| + \sum_{i \in \cI} d_{\cL}(x_i) |\cA_i|$ constraints\footnote{Throughout the paper, when considering LP complexities we will omit constraint complexities due to upper and lower bounds on the LP variables.}, where $d_{\cL}(x_i)$ denotes the number of neighbours of $x_i$ which belong to $\cL$. Note that \eqref{eq:equation-polytope-4} and \eqref{eq:equation-polytope-5} imply that we may view $\bldx \in \cB$ as a random vector, and for each $j \in \cL$ the vector $\bldp_j$ may be interpreted as a probability distribution on the local configuration $\bldx_j \in \cB_j$; \eqref{eq:equation-polytope-theoretical} then expresses each vector $\bldg_i$ (for each $i \in \cI$) as the induced probability distribution on $x_i \in \cA_i$. It may be easily checked that $\boldsymbol{\lambda} \bldg ^T$ is then the expectation of $\log \prod_{i \in \cY} h_i(x_i)$ with respect to this distribution\footnote{If $\bldg$ represents a probability distribution on $\bldx$, we denote the expectation of $F(\bldx)$ with respect to the distribution $\bldg$ as $\mathbb{E}_{\bldgg} \, F(\bldx)$.}; this interpretation of the cost function will be useful in our treatment of system pseudodistance in Section \ref{sec:PCFs}.
A similar probabilistic interpretation was also considered in the context of pseudocodewords of graph-cover decoding in \cite{FKKR}\footnote{Another interpretation of the polytope $\cQ$ is that the projection of $\cQ$ onto $\bldg$ is formed by the intersection of convex hulls corresponding to the local behaviors, i.e., $(\bldg_i)_{i \in \cI_j} \in \mathrm{conv} \{ \bldXi(\cB_j) \}$ for all $j \in \cL$.}.

If the LP solution $(\bar{\bldg}_\mathrm{out}, \bldp)$ is an integral point in $\cQ$ (i.e., all of its coordinates are integers), the receiver output is the configuration $\bldx_\mathrm{out} = \bldP_{\cYY}^{-1} \left( \bar{\bldXi}^{-1} (\bar{\bldg}_\mathrm{out}) \right)$ (we shall prove in the next section that this output is indeed in $\cB$). Of course, in the communications context, we are usually only interested in a subset of the configuration symbols, namely the information bits. If the LP solution is not integral, the receiver reports ${\tt FAILURE}$.

\section{Efficient Linear-Programming Relaxation and its Properties}
\label{sec:efficient_LP}
We next define another linear program, and prove that its performance is equivalent to that defined in Section \ref{sec:efficient_LP_theoretical}. This new program achieves lower description complexity than {\bf LP2} by removing unnecessary constraints from the formulation. We remove constraints in two ways: from the variable set, and by defining constraints with respect to an `anchor node'.

For each $i\in\cI$, let $\alpha_i$ be an arbitrary element of $\cA_i$, and let $\cA_i^{-} = \cA_i \backslash \{ \alpha_i \} $ (note that for each $i\in\cI$, $|\cA_i| \ge 2$, otherwise $x_i$ is not a `variable'). For each $i \in \cI$, $\alpha \in \cA_i$, define $\tilde{\bldxi}_i (\alpha) = ( \mathbb{I}(\gamma = \alpha) )_{\gamma \in A_i^{-} }$. Note that this indicator vector is the same as $\bldxi_i (\alpha)$ except that the entry corresponding to $\alpha_i$ has been removed. Correspondingly, for each $\bldx_{\cYY} \in \cA_{\cYY}$ we define the indicator vector $\tilde{\bldXi}(\bldx_{\cYY}) = ( \tilde{\bldxi}_i(x_i) )_{i \in\cY}$. Again, the mapping $\tilde{\bldXi}$ is injective.

Now, we define the vector $\tilde{\bldg}$ similarly to $\bldg$ but with entries corresponding to each $\alpha_i$ removed, i.e., 
\[
\tilde{\bldg} = ( \tilde{\bldg}_i )_{i \in \cYY} \quad \mbox{where} \quad \tilde{\bldg}_i = ( \tilde{g}_i^{(\alpha)} )_{\alpha \in \cA_i^{-}} \quad \forall i \in \cY \; ,
\] 
and we define the vector $\tilde{\boldsymbol{\lambda}}$ by
\[
\tilde{\boldsymbol{\lambda}} = ( \tilde{\boldsymbol{\lambda}}_i )_{i \in \cYY} \quad \mbox{where} \quad \tilde{\boldsymbol{\lambda}}_i = ( \tilde{\lambda}_i^{(\alpha)} )_{\alpha \in \cA_i^{-}} \quad \forall i \in \cY \; ,
\]  
and $\tilde{\lambda}_i^{(\alpha)} = \log \left[ h_i(\alpha) / h_i(\alpha_i) \right]$ for each $i \in \cY$, $\alpha \in \cA_i^{-}$. 

Also, for each $i\in\cI\backslash \cY$, let $t(i)$ be an arbitrary element of $\cJ_i$, i.e., $f_{t(i)}$ is an arbitrary neighbouring factor node of the non-observable variable $x_i$ and is referred to as the ``anchor node'' for that variable node in the factor graph. The LP is then as follows.

\vspace{1pc}
\hspace{-3mm}\fbox{
\begin{minipage}{0.465\textwidth}
{\bf LP3: Low-complexity Relaxation}

\vspace{1mm}

{\bf Cost Function:} $\tilde{\boldsymbol{\lambda}} \tilde{\bldg} ^T$

{\bf Constraints (Polytope $\tilde{\cQ}$):} Constraints \eqref{eq:equation-polytope-4} and \eqref{eq:equation-polytope-5}, together with
\begin{eqnarray}
& \forall i \in \cY, \; \forall j \in \cJ_i \cap \cL, \; \forall \alpha \in \cA_i^{-}, \nonumber \\
& \tilde{g}_i^{(\alpha)} = 
\sum_{\bldbb \in \cB_j: \, b_i=\alpha} p_{j,\bldbb}
\label{eq:equation-polytope-6} 
\end{eqnarray} 
and
\begin{eqnarray}
& \forall i \in \cI\backslash\cY, \; \forall j \in \cJ_i \backslash \{t(i)\}, \; \forall \alpha \in \cA_i^{-}, \nonumber \\
& \sum_{\bldbb \in \cB_j: \, b_i=\alpha} p_{j,\bldbb} = 
\sum_{\bldbb \in \cB_{t(i)}: \, b_i=\alpha} p_{t(i),\bldbb} \; .
\label{eq:equation-polytope-7} 
\end{eqnarray} 
{\bf Receiver Output:}
\begin{equation}
\left\{ \begin{array}{cc}
\bldx_\mathrm{out} = \tilde{\bldXi}^{-1} (\tilde{\bldg}_\mathrm{out}) \in \cB & \textrm{ if } (\tilde{\bldg}_\mathrm{out}, \bldp) \textrm{ is integral } \\
{\tt FAILURE} & \textrm{ otherwise. }\end{array}\right.
\label{eq:LP3_output} 
\end{equation} 
\end{minipage}}
\vspace{1pc}

The receiver output is equal to the configuration $\bldx_\mathrm{out} = \tilde{\bldXi}^{-1} (\tilde{\bldg}_\mathrm{out}) \in \cB$ in the case where the solution $(\tilde{\bldg}_\mathrm{out}, \bldp)$ to {\bf LP3} is an integral point in $\cQ$, and reports ${\tt FAILURE}$ if the solution is not integral. {\bf LP3} comprises $\sum_{j \in \cL} |\cB_j| + \sum_{i \in \cY} |\cA_i| - |\cY|$ variables and $|\cL| + \sum_{i \in \cYY} d_{\cL}(x_i) (|\cA_i| - 1) + \sum_{i \in \cI \backslash \cYY} ( d(x_i) - 1 ) ( |\cA_i| - 1)$ constraints, where $d(x_i)$ denotes the degree of $x_i$ and $d_{\cL}(x_i)$ denotes the number of neighbours of $x_i$ which belong to $\cL$.

The following theorem ensures the equivalence of the linear programs {\bf LP2} and {\bf LP3}, and also assures the \emph{optimum certificate} property, i.e., if the output of either LP is a configuration, then it is the optimum configuration. In the communications context, the optimum corresponds to the maximum likelihood transmit configuration; thus in this case we have the \emph{maximum likelihood certificate} property.

\medskip
\begin{theorem}\label{prop:LP_equivalence_2}
The two linear programs {\bf LP2} and {\bf LP3} produce the same output (configuration or {\tt FAILURE}). Also, if either LP output is an integral point in the LP polytope, then it corresponds to the optimum configuration, i.e., $\bldx_\mathrm{out} = \bldx_\mathrm{opt}$. 
\end{theorem}
\medskip
\begin{proof}
It is straightforward to show that the mapping
\begin{eqnarray*}
\bldV \; : \; \tilde{\cQ} & \longrightarrow & \cQ \\
(\tilde{\bldg},\bldp) & \mapsto  & (\bar{\bldg},\bldp)
\end{eqnarray*}
defined by
\[
g_i^{(\alpha)} = 
\left\{ \begin{array}{ccc}
\tilde{g}_i^{(\alpha)} & \textrm{ if } i \in \cY, \alpha \in \cA_i^{-} \\
1 - \sum_{\beta \in \cA_i^{-}} \tilde{g}_i^{(\beta)} & \textrm{ if }  i \in \cY, \alpha = \alpha_i \\
\sum_{\bldbb \in \cB_{t(i)}: \, b_i=\alpha} p_{{t(i)},\bldbb}  & \textrm{ if } i \in \cI \backslash \cY \end{array}\right.
\]
and with inverse 
\[
\tilde{g}_i^{(\alpha)} = g_i^{(\alpha)} \quad \forall i \in \cY, \alpha \in \cA_i^{-}
\]
is a bijection from one polytope to the other (i.e., $\tilde{\bldg}$ satisfies the constraints of {\bf LP3} for some vector $\bldp$ if and only if $\bar{\bldg}$ with $(\bar{\bldg}, \bldp) = \bldV(\tilde{\bldg}, \bldp)$ satisfies the constraints of {\bf LP2} for the same vector $\bldp$). Also
\begin{eqnarray}
\tilde{\boldsymbol{\lambda}} \tilde{\bldg} ^T & = & \sum_{i \in \cI} \sum_{\alpha \in \cA_i^{-}} \left( \log h_i(\alpha) - \log h_i(\alpha_i) \right) \tilde{g}_i^{(\alpha)} \nonumber \\
& = & \sum_{i \in \cI} \left( \sum_{\alpha \in \cA_i^{-}} \log h_i(\alpha) g_i^{(\alpha)} - \log h_i(\alpha_i) [ 1 - g_i^{(\alpha_i)} ] \right) \nonumber \\
& = & \boldsymbol{\lambda} \bldg ^T - \sum_{i \in \cI} \log h_i(\alpha_i) \; ,
\label{eq:cost_fn_equivalence}
\end{eqnarray} 
implying that the bijection $\bldV$ preserves the cost function up to an additive constant. 

Next, we prove that for every configuration $\bldx \in \cB$, there exists $\bldp$ such that $(\bar{\bldXi} \left(\bldx \right), \bldp) \in \cQ$. Let $\bldx \in \cB$, and define
\[
\forall j \in \cL, \bldb \in \cB_j, \quad p_{j,\bldbb} = 
\left\{ \begin{array}{cc}
1 & \textrm{ if } \bldb = (x_i)_{i \in \cI_j} \\
0 & \textrm{ otherwise. } \end{array}\right.
\]
Letting $\tilde{\bldg} = \tilde{\bldXi}(\cP_{\cYY}(\bldx))$ and $\bar{\bldg} = \bar{\bldXi}(\bldx)$, it is easy to check that $(\tilde{\bldg}, \bldp) \in \tilde{\cQ}$ and $(\bar{\bldg}, \bldp) \in \cQ$ (and that in fact $(\bar{\bldg},\bldp) = \bldV(\tilde{\bldg}, \bldp)$). This property ensures that every valid configuration $\bldx \in \cB$ has a ``representative" in the polytope, and thus is a candidate for being output by the receiver.

Next, let $(\tilde{\bldg}, \bldp) \in \tilde{\cQ}$ and let $\bar{\bldg}$ be such that $(\bar{\bldg}, \bldp) = \bldV(\tilde{\bldg}, \bldp) \in \cQ$. Suppose that all of the coordinates of $\bldp$ are integers. Then, by (\ref{eq:equation-polytope-4}) and (\ref{eq:equation-polytope-5}), for any $j \in \cL$ we must have 
\[
\forall \bldb \in \cB_j, \quad p_{j,\bldbb} = 
\left\{ \begin{array}{cc}
1 & \textrm{ if } \bldb = \bldb^{(j)} \\
0 & \textrm{ otherwise } \end{array}\right.
\]
for some $\bldb^{(j)} \in \cB_j$.

Next we note that for any $i \in \cI$, $j,k \in \cJ_i \cap \cL$, if $b^{(j)}_i = \alpha$ then (using (\ref{eq:equation-polytope-theoretical}))
\begin{equation}
g_i^{(\alpha)} = \sum_{\bldbb \in \cB_j: \, b_i=\alpha} p_{j,\bldbb} = 1 = \sum_{\bldbb \in \cB_k: \, b_i=\alpha} p_{k,\bldbb}
\label{eq:starstar}
\end{equation}
and thus $b^{(k)}_i = \alpha$. Therefore, there exists $\bldx \in \cA$ such that 
\[
(x_i)_{i \in \cI_j} = \bldb^{(j)} \quad \forall j \in \cL \; .
\]
Therefore, $\bldx$ is a valid configuration ($\bldx \in \cB$). Also we may conclude from (\ref{eq:starstar}) that
\[
g_i^{(\alpha)} = 
\left\{ \begin{array}{cc}
1 & \textrm{ if } x_i = \alpha \\
0 & \textrm{ otherwise } \end{array}\right.
\] 
and therefore $\bar{\bldg} = \bar{\bldXi} \left( \bldx \right)$. Also, from the definition of the mapping $\bldV$, we have $\tilde{\bldg} = \tilde{\bldXi} \left( \bldP_{\cYY}\left(\bldx\right) \right)$.

\begin{figure}[t]
\begin{center}\includegraphics[%
  width=\columnwidth, keepaspectratio]{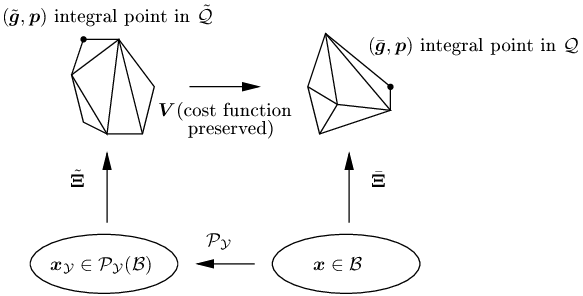}\end{center}
\caption{Illustration of the relationships involved in the proof of Theorem \ref{prop:LP_equivalence_2}.\label{cap:mappings}}
\end{figure}

Summarizing these results, we conclude that $(\bar{\bldg}_\mathrm{opt}, \bldp) \in \cQ$ optimizes the cost function $\boldsymbol{\lambda} \bldg ^T$ over $\cQ$ and is integral if and only if $(\tilde{\bldg}_\mathrm{opt}, \bldp) = \bldV^{-1}(\bar{\bldg}_\mathrm{opt}, \bldp) \in \tilde{\cQ}$ optimizes the cost function $\tilde{\boldsymbol{\lambda}} \tilde{\bldg} ^T$ over $\tilde{\cQ}$ and is integral, where $\bar{\bldXi}^{-1}(\bar{\bldg}) = \bldx \in \cB$ and $\bldx_{\cYY} = \tilde{\bldXi}^{-1}(\tilde{\bldg}) = \cP_{\cYY}(\bldx)$. A graphical illustration of these relationships is shown in Figure \ref{cap:mappings}. 
\end{proof}
\medskip
Thus both LP receivers output either the optimum configuration or ${\tt FAILURE}$, and have the same performance. {\bf LP3} has lower descriptive complexity and is suitable for implementation (we shall use it to solve the joint equalization and decoding problem in Section \ref{sec:equalization_decoding}); however, for theoretical work {\bf LP2} is more suitable (we shall use this LP throughout Section \ref{sec:PCFs}). 

\section{Pseudoconfigurations} 
\label{sec:PCFs}
In this section, we show a connection between the failure of the LP and SPA receivers based on \emph{pseudoconfiguration} concepts, and define a general concept of \emph{pseudodistance} for LP receivers. 

\subsection{Connecting the failure mechanisms of the LP and SPA receivers}  

We first define what is meant by a finite cover of a factor graph.  

\medskip
\begin{definition}
Let $M$ be a positive integer, and let $\cM = \{ 1, 2, \cdots, M \}$. Let $\cG$ be the factor graph corresponding to the global function $u$ and its factorization given in (\ref{eq:factorization_of_global_function}). A \emph{cover configuration} of degree $M$ is a vector 
$\bldx^{(M)} = ( \bldx_i^{(M)} )_{i \in \cI}$ where $\bldx_i^{(M)} = (x_{i,m})_{m \in \cM} \in \cA_i^M$ for each $i \in \cI$.
Define $u^{(M)}$ as the following function of the cover configuration $\bldx^{(M)}$ of degree $M$:   
\begin{equation}
u^{(M)}\left(\bldx^{(M)}\right)=\prod_{m \in \cM} \prod_{j\in \cJ}f_{j}\left(\bldx_{j,m}\right)
\label{eq:cover_graph_factorization}
\end{equation}
where, for each $j\in\cJ$, $i\in\cI_j$, $\Pi_{j,i}$ is a permutation on the set $\cM$, and for each $j\in\cJ$, $m\in\cM$,
\[
\bldx_{j,m} = ( x_{i,\Pi_{j,i}(m)} )_{i \in\cI_j} \; .
\]
A \emph{cover} of the factor graph $\cG$, of degree $M$, is a factor graph for the global function $u^{(M)}$ and its factorization (\ref{eq:cover_graph_factorization}). In order to distinguish between different factor node labels, we write (\ref{eq:cover_graph_factorization}) as
\[
u^{(M)}\left(\bldx^{(M)}\right)=\prod_{m\in \cM} \prod_{j\in \cJ}f_{j,m}\left(\bldx_{j,m}\right)
\]
where $f_{j,m} = f_{j}$ for each $j\in\cJ$, $m\in\cM$.
\end{definition}
\medskip

It may be seen that a cover graph of degree $M$ is a graph whose vertex set consists of $M$ copies of $x_i$ (labeled $x_{i,m}$) and $M$ copies of $f_j$ (labeled $f_{j,m}$), such that for each $j\in\cJ$, $i\in\cI_j$, the $M$ copies of $x_i$ and the $M$ copies of $f_j$ are connected in a one-to-one fashion determined by the permutation $\Pi_{j,i}$.

\medskip
\begin{definition}
The \emph{cover behavior} $\cB_M$ is defined as the set of all cover configurations $\bldx^{(M)}$ such that $\bldx_{j,m} \in \cB_j$ for each $j\in\cJ$, $m\in\cM$. For any $M\ge 1$, a \emph{graph-cover pseudoconfiguration} is defined to be a valid cover configuration (i.e., one which lies in the behavior $\cB_M$). 
\end{definition}
\medskip

\begin{definition}
For any graph-cover pseudoconfiguration, the (unscaled) \emph{graph-cover pseudoconfiguration vector} $\bar{\bldeta}$ is defined by 
\[
\bar{\bldeta} = ( \bldeta_i )_{i \in \cI} \quad \mbox{where} \quad \bldeta_i = ( \eta_i^{(\alpha)} )_{\alpha \in \cA_i} \quad \forall i\in\cI
\]
and
\[
\eta_i^{(\alpha)} = \left| \{ m \in \cM : \, x_{i,m} = \alpha \} \right| 
\]
for each $i\in\cI$, $\alpha\in\cA_i$. The \emph{normalized graph-cover pseudoconfiguration vector} $\bar{\bldg}$ is then defined by $\bar{\bldg} = \bar{\bldeta} / M$.
\end{definition}
\medskip

The set of graph-cover pseudocodewords has previously been shown to be responsible, to an approximate degree, for the failure of SPA decoding of binary linear codes (see e.g. \cite{KV-IEEE-IT}). Such arguments generalize in a straightforward manner to the present context; the following provides a brief overview. The SPA receiver passes messages on the edges on the factor graph $\cG$ of the global function; SPA processing begins by passing the message $f_j(x_i)$ from each degree-$1$ factor node $f_j$ neighbouring $x_i$ ($i \in \cYY$), and thereafter follows a preset (usually periodic) message-passing schedule. The message-passing algorithm is ``local'' in that the message passed from any node $a$ to any other node $b$ is a function only of the messages incoming at $a$ from all neighbours of $a$ other than $b$. Assume that the SPA receiver running on the original graph $\cG$ yields the optimum configuration $\bldx_\mathrm{opt}$; recall that this is the maximum, over all valid configurations $\bldx \in \cB$, of the function $\sum_{i \in \cYY} \log h_i(x_i)$. Of course, the SPA receiver does not actually seek to maximize this function, but instead seeks to \emph{marginalize} this function with respect to each relevant local variable $x_i$, and subsequently choose the value of $x_i$ which maximizes each marginal (see \cite{Kschischang} for further details).

Next consider the SPA decoding algorithm operating on a cover graph of $\cG$ of degree $M$, with the same schedule except that we replace message-passing between any pair of original nodes in $\cG$ at any iteration $t$ by \emph{parallel} message-passing between the set of \emph{copies} of these nodes in the cover of $\cG$ at iteration $t$. Then, since SPA processing on the cover graph begins by passing the (replicated) message $f_j(x_{i,m})$ from the (replicated) degree-$1$ factor node $f_{j,m}$ neighbouring $x_{i,m}$, $m \in \cM$ (for each $i \in \cYY$), and the schedule matches at each iteration as described above, a straightforward inductive argument shows that the set of messages passed from nodes $x_{i,m}$ to nodes $f_{j,m}$ (for fixed $i,j$ and considering all $m \in \cM$) in the cover graph at iteration $t$ consist of $M$ identical copies of the message passed from node $x_{i}$ to node $f_{j}$ at iteration $t$. Then the SPA decoder running on the cover graph of degree $M$ of $\cG$ must yield the cover configuration $\bldx^{(M)}$ with $x_{j,m} = x_j$ for all $m \in \cM$ (sometimes called a \emph{lifting} of the configuration \cite{KV-IEEE-IT}). However, if we assume that the SPA receiver returns the maximum, over all valid configurations $\bldx^{(M)} \in \cB^{(M)}$, of the global function, i.e.,
\begin{equation}
\sum_{i \in \cYY} \sum_{m \in \cM} \log h_i(x_{i,m}) = \sum_{i \in \cYY} \sum_{\alpha \in \cA_i} \eta_i^{(\alpha)} \log h_i(\alpha) = \bldlambda^T \bldeta \; ,
\end{equation}
this yields a contradiction whenever there exists a graph-cover pseudoconfiguration with lower cost $\bldlambda^T \bldeta$. 

Note that the above reasoning holds under the assumption that the SPA algorithm has the property that it always returns the optimum configuration for the graph on which it operates. This is only an approximation, and it is for this reason that the role of graph-cover pseudoconfigurations in SPA decoding is only an \emph{approximate} model, whereas for LP receiver the model is \emph{exact}. However the approximation can be quite accurate; SPA decoding failure is \emph{exactly} characterized by the \emph{computation tree pseudocodewords} of the system's factor graph, and the graph-cover pseudocodewords may be taken as an approximation of the computation tree pseudocodewords, since the local neighbourhood of any variable node is identical to some depth in both graphs. For more 
discussion on these connections in the context of linear codes, see e.g. \cite{KV-IEEE-IT}.

\begin{figure*}[t]
\normalsize
\setcounter{mytempeqncnt}{\value{equation}}
\setcounter{equation}{18}
\begin{eqnarray}
d_{\mathrm{eff}}(\bar{\bldx},\bldkappa) = \frac{\mathbb{E}_{\bldgg} \, \| \blds(\bldx_{\cY}) - \blds(\bar{\bldx}_{\cY}) \|^2}{\| \mathbb{E}_{\bldgg} \, \blds(\bldx_{\cY}) - \blds(\bar{\bldx}_{\cY}) \|} 
= \frac{ \left| \displaystyle\sum_{i \in \cY} \left( |s_i(\bar{x}_i)|^2 + \sum_{\alpha \in \cA_i} \left( |s_i(\alpha)|^2 - 2 a_i(\alpha) a_i(\bar{x}_i) - 2 b_i(\alpha) b_i(\bar{x}_i) \right) g_i^{(\alpha)} \right) \right|  }{\sqrt{\displaystyle\sum_{i \in \cY} \left[ \left( \displaystyle\sum_{\alpha \in \cA_i} a_i(\alpha) g_i^{(\alpha)} - a_i(\bar{x}_i) \right)^2 + \left( \displaystyle\sum_{\alpha \in \cA_i} b_i(\alpha) g_i^{(\alpha)} - b_i(\bar{x}_i) \right)^2  \right]}} \; .
\label{eq:pseudodistance}
\end{eqnarray}
\setcounter{equation}{17}
\hrulefill
\vspace*{4pt}
\end{figure*}

\medskip
\begin{definition}
A \emph{linear-programming pseudoconfiguration} (LP pseudoconfiguration) is a rational point $(\bar{\bldg}, \bldp)$ in the polytope $\cQ$ of the linear program {\bf LP2}.
\end{definition}
\medskip

Next, we state the equivalence between the set of LP pseudoconfigurations and the set of graph-cover pseudoconfigurations. The result is summarized in the following theorem. 

\medskip
\begin{theorem}\label{thm:PCW_equivalence}
There exists an LP pseudoconfiguration $(\bar{\bldg}, \bldp)$ if and only if there 
exists a graph-cover pseudoconfiguration with normalized pseudoconfiguration vector $\bar{\bldg}$. 
\end{theorem}
\medskip

The proof of Theorem \ref{thm:PCW_equivalence} follows the lines of the proof of \cite[Theorem 7.1]{FSBG_journal}; the details are omitted. Theorem \ref{thm:PCW_equivalence} shows that the pseudoconfigurations which \emph{exactly} characterize the performance of the LP receiver are precisely equivalent to the pseudoconfigurations which (due to the argument above) \emph{approximately} characterize performance of the SPA receiver.

\subsection{Pseudodistance}
\label{pseudodistance}

In this section we define the concept of \emph{system pseudodistance} for communication systems where the set of variables $\{x_i\}_{i \in \cY}$ is observed through complex additive white Gaussian noise (AWGN). For each $i \in \cY$, we have an observation $y_i = p_i + \imath q_i$ which is formed by passing the symbol $x_i$ through a modulation mapper and adding complex Gaussian noise with variance $\sigma^2$ per real dimension (here $\imath = \sqrt{-1}$). The mapper operates according to the following rule: for $i \in \cY$, $\alpha \in \cA_i$ is mapped to $s_i(\alpha) = a_i(\alpha) + \imath b_i(\alpha)$. Note that this  includes cases where different symbols may use different mappers, e.g. orthogonal frequency division multiplexing (OFDM) systems with adaptive modulation. Then
\begin{equation}
h_i(\alpha) = p(y_i | \alpha) = \frac{1}{2 \pi \sigma^2} \exp \left( - \frac{ \left| y_i-s_i(\alpha) \right|^2}{2 \sigma^2} \right)
\label{eq:Gaussian}
\end{equation}
In what follows, we denote the transmitted and received vectors by $\blds(\bldx_{\cY}) = (s_i(x_i))_{i \in \cY}$ and $\bldy = (y_i)_{i \in \cY}$ respectively.

Suppose that the actual transmitter configuration is $\bar{\bldx} \in \cB$, and let $\bldw = \bldXi(\cP_{\cY}(\bar{\bldx}))$. The LP receiver {\bf LP2} favours the pseudoconfiguration $\bldkappa = (\bar{\bldg},\bldp) \in \cQ$ over $\bar{\bldx}$ if and only if $\bldlambda \bldg ^T > \bldlambda \bldw ^T$, i.e., if and only if
\[
\mathbb{E}_{\bldgg} \, \log \prod_{i \in \cY} h_i(x_i) > \mathbb{E}_{\bldww} \, \log \prod_{i \in \cY} h_i(x_i)
\]
Using \eqref{eq:Gaussian}, this condition is easily seen to be equivalent to
\[
\mathbb{E}_{\bldgg} \, \| \bldy - \blds(\bldx_{\cY}) \|^2 < \mathbb{E}_{\bldww} \, \| \bldy - \blds(\bldx_{\cY}) \|^2 = \| \bldy - \blds(\bar{\bldx}_{\cY}) \|^2 \; .
\]
Using $\mathbb{E}_{\bldgg} \, \| \bldy \|^2 = \| \bldy \|^2$, this may be rewritten as
\[
\sum_{i \in \cY} \left( M_i p_i + N_i q_i \right) > R \; ,
\]
where we introduce $M_i = 2\left( \mathbb{E}_{\bldgg} \, a_i(x_i) - a_i(\bar{x}_i) \right)$, $N_i = 2\left( \mathbb{E}_{\bldgg} \, b_i(x_i) - b_i(\bar{x}_i) \right)$, and 
\[
R = \mathbb{E}_{\bldgg} \, \| \blds(\bldx_{\cY}) \|^2 - \| \blds(\bar{\bldx}_{\cY}) \|^2 \; .
\]
In the absence of noise, the modulated signal point in the signal space with $2|\cY|$ dimensions and coordinates $\{p_i\}_{i \in \cY}$ and $\{q_i\}_{i \in \cY}$ is given by $p_i = a_i(\bar{x}_i)$ and $q_i = b_i(\bar{x}_i)$ for all $i \in \cY$. The squared Euclidean distance from this point to the plane $\sum_{i \in \cY} (M_i p_i + N_i q_i) = R$ is then given by $D^2 = (R-S)^2/V$, where
\[
S = 2 \sum_{i \in \cY} \left[ a_i(\bar{x}_i) \mathbb{E}_{\bldgg} \, a_i(x_i) + b_i(\bar{x}_i) \mathbb{E}_{\bldgg} \, b_i(x_i) \right] -2 \| \blds(\bar{\bldx}_{\cY}) \|^2  
\]
and
\[
V = 4 \| \mathbb{E}_{\bldgg} \, \blds(\bldx_{\cY}) - \blds(\bar{\bldx}_{\cY}) \|^2 \; .
\]
Thus the decision boundary is the same as that induced under ML reception by a signal vector at a Euclidean distance $2D$ from the transmit signal vector in the signal space; this motivates the following definition.
\medskip
\begin{definition}
The \emph{effective Euclidean distance} or \emph{system pseudodistance} $d_{\mathrm{eff}}(\bar{\bldx},\bldkappa)$ between the configuration $\bar{\bldx} \in \cB$ and the pseudoconfiguration $\bldkappa = (\bar{\bldg},\bldp) \in \cQ$ is given by (19) at the top of the page.
\setcounter{equation}{19}
\end{definition}
\medskip
This generalizes the concept of pseudodistance given in \cite{FKKR} and \cite{Kelley-Sridhara-ISIT-2006} for binary and nonbinary codes, and in particular generalizes \cite[Theorem 2.1]{FKKR} which was proved for real AWGN and pseudocodewords of a balanced computation tree for a nonbinary code. Note that the system pseudodistance depends on the transmitter configuration (i.e., the information word); while it was proved in \cite{Feldman} for binary codes and in \cite{Flanagan_cw_ind} for nonbinary codes that under a sufficent channel symmetry condition the performance of {\bf LP2} is independent of the codeword transmitted, this property does not hold in the current more general context. The pairwise error probability between the transmitter configuration $\bldx \in \cB$ and the pseudoconfiguration $\bldkappa \in \cQ$ is given by 
\begin{equation}
P_e(\bldx,\bldkappa) = Q \left( \frac{d_{\mathrm{eff}}(\bldx,\bldkappa)}{2 \sigma} \right)
\label{eq:prob_of_pcf_error}
\end{equation}
where $Q(z) = \frac{1}{2\pi} \int_z^{\infty} \exp (-t^2/2) \; \mathrm{d}t$ is the Gaussian $Q$-function. We define the \emph{minimum pseudodistance} of the system as
\[
d_{\mathrm{eff}}^{\mathrm{min}} = \min_{\bldxx \in \cB,\bldsubkappa \in \cQ(\bldxx)} d_{\mathrm{eff}}(\bldx,\bldkappa)
\]
where $\cQ(\bldx)$ denotes the polytope $\cQ$ with the pseudoconfiguration corresponding to $\bldx$ removed. The minimum pseudodistance provides an important single parameter with which to measure system performance, as it plays an analogous role in the context of LP reception over AWGN to that played by the minimum distance of binary linear codes in the context of ML decoding over AWGN. Note that in most cases of practical interest, there are $2^k$ equiprobable transmit configurations $\bldx \in \cB$; therefore the FER performance may be bounded at any SNR according to
\begin{equation}
\frac{1}{2^k} \! \sum_{\bldxx \in \cB} Q \left( \frac{d_{\mathrm{eff}}^{\mathrm{min}} (\bldx)}{2 \sigma} \right) \! \le \! \mbox{FER} \! \le \! \frac{1}{2^k} \! \sum_{\bldxx \in \cB} \sum_{\bldsubkappa \in \cQ(\bldxx)} \!\! Q \left( \frac{d_{\mathrm{eff}} (\bldx, \bldkappa)}{2 \sigma} \right) 
\label{eq:QQQ}
\end{equation}
where $d_{\mathrm{eff}}^{\mathrm{min}} (\bldx) = \min_{\bldsubkappa \in \cQ(\bldxx)} d_{\mathrm{eff}} (\bldx, \bldkappa)$. Note also that pseudoconfigurations at $d_{\mathrm{eff}}^{\mathrm{min}}$ begin to dominate the right-hand side (union bound) expression in \eqref{eq:QQQ} at sufficiently high SNR. 

Finally, it is straightforward to show that in the case where the pseudoconfiguration $\bldkappa \in \cQ$ corresponds to a configuration $\bldz \in \cB$, we have $\bldg = \bldXi(\cP_{\cY}(\bldz))$ and (\ref{eq:pseudodistance}) reduces to 
\[
d_{\mathrm{eff}}(\bar{\bldx},\bldkappa) = \| \blds(\bldP_{\cY}(\bldz)) - \blds(\bar{\bldx}_{\cY}) \|
\]
which is the ordinary Euclidean distance between the two relevant modulated signals.

For the case of real AWGN with variance $\sigma^2$ per dimension (here $y_i$ and $s_i(\alpha)$ are real for each $i \in \cY$), a similar analysis shows that assuming the transmitter configuration is $\bar{\bldx} \in \cB$, the probability of error due to pseudoconfiguration $\bldkappa = (\bar{\bldg},\bldp) \in \cQ$ is again given by (\ref{eq:prob_of_pcf_error}), where
\begin{align}
& d_{\mathrm{eff}}(\bar{\bldx},\bldkappa) = \nonumber \\
& \frac{ \left| \displaystyle\sum_{i \in \cY} \left( s_i^2(\bar{x}_i) + \displaystyle\sum_{\alpha \in \cA_i} \left( s_i^2(\alpha) - 2 s_i(\alpha) s_i(\bar{x}_i) \right) g_i^{(\alpha)} \right) \right| }{\sqrt{\displaystyle\sum_{i \in \cY} \left( \displaystyle\sum_{\alpha \in \cA_i} s_i(\alpha) g_i^{(\alpha)} - s_i(\bar{x}_i) \right)^2 }} \nonumber \\
& = \frac{ \left| \sum_{i \in \cY} \left( t_i^2 + v_i - 2 t_i m_i \right) \right| }{\sqrt{\sum_{i \in \cY} \left( m_i - t_i \right)^2 }} \label{eq:pseudodistance_real}
\end{align}
where we define $\bldt = \left( t_i \right)_{i \in \cY}$, $\bldm = \left( m_i \right)_{i \in \cY}$ and $\bldv = \left( v_i \right)_{i \in \cY}$, and for each $i \in \cY$ we have $t_i = s_i(\bar{x}_i)$,
\[
m_i = \mathbb{E}_{\bldgg} \, s_i(x_i) = \sum_{\alpha \in \cA_i} s_i(\alpha) g_i^{(\alpha)}
\]
and 
\[
v_i = \mathbb{E}_{\bldgg} \, s_i^2(x_i) = \sum_{\alpha \in \cA_i} s_i^2(\alpha) g_i^{(\alpha)} \; .
\]

\section{Example Application: LP-Based Joint Equalization and Decoding} 
\label{sec:equalization_decoding}

In this section we consider an example application where we use the above framework to design an LP receiver for a system using linear coding and memoryless modulation over a frequency selective channel with AWGN. 

\subsection{System model and notation}
\label{system_model_and_notation}
The system model may be described as follows. Information-bearing data are encoded to form codewords of the (binary or nonbinary) code $\code$ over the ring $\rrr$, characterized by the $m \times n$ \emph{parity-check matrix} $\cH = \left( H_{j,i} \right)$ over $\rrr$. Denote the set of coded symbol indices and parity-check indices by $\cU = \{ 1,2,\cdots, n\}$ and $\cV = \{ 1,2,\cdots, m\}$ respectively. For each $j \in \cV$, define the $j$-th \emph{local code} over $\rrr$ by 
\[
\code_j = \{ (c_i)_{i \in \cU_j} : \, \sum_{i \in \cU_j} H_{j,i} c_i = 0 \}
\]
where $\cU_j \subseteq \cU$ is the support of the $j$-th row of $\cH$ for each $j \in \cV$, and multiplication and addition are over $\rrr$. Thus $\bldc \in \code$ if and only if $\bldc_j \triangleq (c_i)_{i \in \cU_j}$ lies in $\code_j$ for each $j \in \cV$. 

Each coded symbol $c_i$ is mapped directly to a modulation symbol $x_i = X(c_i) \in \cT$, where $\cT \subset \mathbb{C}$ denotes the transmit constellation. The (injective) modulation mapping is defined by $X \; : \; \rrr \rightarrow \cT$. The modulated symbols are transmitted over a (possibly time-variant) frequency selective channel with AWGN; the received signal is given by
\[
y_i = \sum_{t=0}^{L} h_t^{(i)} x_{i-t} + n_i
\]
where $n_i$ is a zero-mean complex Gaussian random variable with variance $\sigma^2$. We assume that the receiver has complete knowledge of the set of complex channel coefficients $\{ h_t^{(i)} \}$.

We adopt a state-space (trellis) representation for the channel with state space $\cS = {\rrr}^L$; also let $\cS^{-} = \cS \backslash \{ \bldzeros_{1 \times L} \}$. The local behavior (trellis edge set) for the state-space model is denoted by $\cD$. For $\bldd \in \cD$, let $\ip(\bldd)$, $\op_i(\bldd)$, $s^{S}(\bldd)$ and $s^{E}(\bldd)$ denote the channel input, output (at time index $i$), initial state and final state respectively. Thus if we set $\cD = {\rrr}^{L+1}$ and adopt the notation $\bldd = (d_0 \; d_1 \; \cdots \; d_L) \in \cD$, we may have $\ip(\bldd) = d_0$, $s^S(\bldd) = (d_1 \; d_2 \; \cdots \; d_L)$, $s^E(\bldd) = (d_0 \; d_1 \; \cdots \; d_{L-1})$, and $\op_i(\bldd) = \sum_{t=0}^{L} h_t^{(i)} X(d_t)$. Also let $\cD^{-} = \cD \backslash \{ \bldzeros_{1 \times (L+1)} \}$. 

Finally, we note that the common case where $\rrr$ is a finite field is included as a special case of this framework, and also that this system is a generalization of the system of \cite{FSBG_journal} to frequency selective channels.

\subsection{Factor graph and linear-programming receiver}
\label{factor_graph_and_LP_rx}
We next derive the factor graph for the communication problem. Denote the state sequence followed by the channel by $\blds = (s_0 \: s_1 \: \cdots \: s_n)$, and the corresponding sequence of trellis edges by $\bldd = (\bldd_1 \: \bldd_2 \: \cdots \: \bldd_n)$, where $\bldd_i = (c_i \: c_{i-1} \: \cdots \: c_{i-L})$ for $i \in \cU$. For the purpose of exposition we assume that the final channel state $s_n$ is unknown to the receiver, but that the initial channel state $s_0$ is known to the receiver and is $\bldzeros_{1 \times L}$. We assume that each codeword in $\code$ is transmitted with equal probability. Using Bayes' rule, the \emph{a posteriori} probability of the transmitter-channel configuration conditioned on the entire received data is given by (here $P$ denotes probability, and $p$ denotes probability density) 
\[
P(\bldc,\blds,\bldd | \bldy) = \frac{p(\bldy | \bldd) P(\bldc,\blds,\bldd)}{p(\bldy)}
\]
Thus the global function is given by\footnote{For a system with transmitter-channel configurations $\bldx \in \cX$ and received observations $\bldy$, setting $u(\bldx) = K \cdot P(\bldx | \bldy)$ (where $K$ does not depend on $\bldx$) implies that the receiver decision rule $\hat{\bldx} = \arg \max_{\bldxx \in \cA} u(\bldx)$ minimizes the configuration error probability.}
\begin{equation}
u(\bldc,\blds,\bldd) \!=\! \prod_{i \in \cU} Q_i(\bldd_i) \prod_{j \in \cV} \chi_j(\bldc_j) \prod_{i \in \cU} T_i(c_i,\bldd_i,s_{i-1},s_i) \nu(s_0)
\label{eq:factorization_turbo_eq}
\end{equation}
Here $Q_i(\bldd_i) = p(y_i | \bldd_i)$ for each $i \in \cU$, and $\chi_j(\bldc_j) = \mathbb{I}(\bldc_j \in \code_j)$ for each $j \in \cV$. The factor $T_i$ for each $i\in \cU$ represents the channel state-space constraint and may be written as $T_i(c,\bldd,s^{(I)},s^{(F)}) = \mathbb{I}(c = d_0) \cdot \mathbb{I}(s^{(I)} = (d_1 \; d_2 \; \cdots \; d_{L})) \cdot \mathbb{I}(s^{(F)} = (d_0 \; d_1 \; \cdots \; d_{L-1}))$. The factor $\nu(s_0) = \mathbb{I}(s_0=\bldzeros_{1 \times L})$ expresses the receiver's knowledge of the initial state of the channel. 

\begin{figure}[t]
\begin{center}\includegraphics[%
  width=1.0\columnwidth, keepaspectratio]{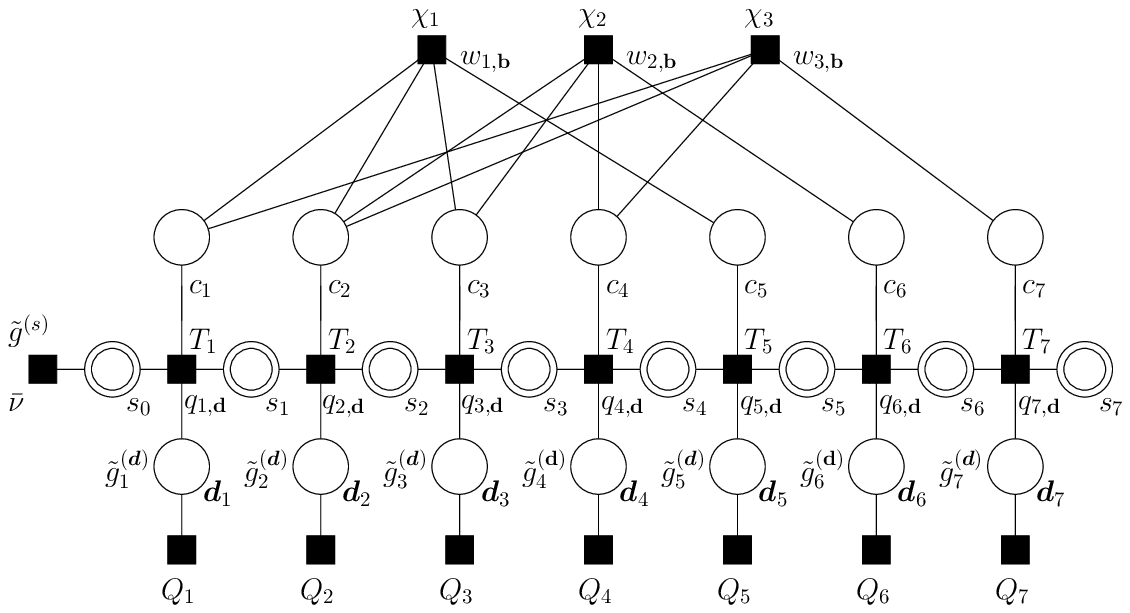}\end{center}
\caption{Factor graph for coded transmissions with memoryless modulation over a frequency selective channel. The factor graph is illustrated for $n=7$, $m=3$, and the binary $[7,4]$ Hamming code. Here $\cY = \{ \bldd_1,\bldd_2,\cdots,\bldd_7 \}$ and $\cL = \{ \bar{\nu}, \chi_1,\chi_2,\chi_3,T_1,T_2,\cdots,T_7\}$. Also indicated on the graph are the relevant LP variables. The constraints of the LP (acting on these variables) may be read directly from the edges of the factor graph.\label{cap:factor_graph}}
\end{figure}

The factor graph corresponding to the global function and its factorization given by (\ref{eq:factorization_turbo_eq}) is illustrated in Figure \ref{cap:factor_graph} for $n=7$, $m=3$, and the binary $[7,4]$ Hamming code. Here the set of indicator function type factor nodes is $\cL = \{ \bar{\nu}, \chi_1,\chi_2,\chi_3,T_1,T_2,\cdots,T_7\}$, and the set of ``observable'' variable nodes is $\cY = \{ \bldd_1,\bldd_2,\cdots,\bldd_7 \}$. 

The LP is then derived using the rules defined in Section \ref{sec:efficient_LP}. After some simplifications\footnote{Consisting primarily of the elimination of variables $\tilde{g}^{(s)}$ and $\tilde{g}_1^{(\blddd)}$.}, this reduces to the following; each constraint is marked with the corresponding constraint from {\bf LP3} from which it derives. Here $\cU^{-} = \cU\backslash\{n\}$ and $\rrr^{-} = \rrr\backslash\{0\}$. Also, for each $i \in \cU$, $T_i$ acts as anchor node for $c_i$, and $q_{i,d}$ acts as anchor node for $s_{i-1}$.

\vspace{1pc}
\hspace{-3mm}\fbox{
\begin{minipage}{0.465\textwidth}
{\bf LP4: Joint Equalization and Decoding}

\vspace{1mm}

{\bf Cost Function:} 
\begin{equation}
\sum_{i \in \cU} \sum_{\blddd \in\cD^{-}} \tilde{\lambda}_i^{(\blddd)} q_{i,\blddd}
\label{eq:cost_fn_turbo_eq_1}
\end{equation}
where we have, for $i\in\cU$, $\bldd \in\cD^{-}$, 
\begin{equation}
\tilde{\lambda}_i^{(\blddd)} \!=\! \log \left( \frac{Q_i(\bldd)}{Q_i(\zeros)} \right) \!=\! \frac{ \left( \left| y_i - \op_i(\zeros) \right|^2 \!-\! \left| y_i - \op_i(\bldd) \right|^2 \right)}{\sigma^2} \; .
\end{equation}
{\bf Constraints (Polytope $\tilde{\cQ}$):} 
\begin{equation} 
\forall j \in \cV, \; \forall \bldb \in \code_j,  w_{j,\bldbb} \ge 0 \; ; \: \forall i \in \cU, \; \forall \bldd \in \cD, q_{i,\blddd} \ge 0 
\end{equation} 
from (\ref{eq:equation-polytope-4}), 
\begin{equation}
\forall j \in \cV, \quad \sum_{\bldbb \in \code_j} w_{j,\bldbb} = 1 \; ; \quad \forall i \in \cU, \quad \sum_{\blddd \in \cD} q_{i,\blddd} = 1
\label{eq:sum_w}
\end{equation}
from (\ref{eq:equation-polytope-5}), 
\begin{equation}
\forall i \in \cU^{-}, \forall s \in \cS^{-}, \!\! \sum_{\blddd \in \cD: \, s^{E}(\blddd)=s} q_{i,\blddd} = \!\! \sum_{\blddd \in \cD: \, s^{S}(\blddd)=s} q_{i+1,\blddd} \; ,
\label{eq:sum_q_to_next_q}
\end{equation}
together with
\begin{equation}
\forall i \in \cU, \forall j \in \cU_j, \forall r \in \rrr^{-}, \!\! \sum_{\blddd \in \cD: \, \ip(\blddd)=r} q_{i,\blddd} = \!\! \sum_{\bldbb \in \code_j: \, b_i=r} w_{j,\bldbb}
\label{eq:sum_q_to_w}
\end{equation}
and 
\begin{equation}
\forall s \in \cS^{-}, \quad \sum_{\blddd \in \cD: \, s^{S}(\blddd)=s} q_{1,\blddd} = 0
\label{eq:sum_q_to_gtilde}
\end{equation}
from (\ref{eq:equation-polytope-7}).

{\bf Receiver Output:}
Set $f_i^{(r)} = \sum_{\blddd \in \cD: \, \ip(\blddd)=r} q_{i,\blddd}$ for each $r \in \rrr^{-}$. Then the receiver output is
\begin{equation}
\left\{ \begin{array}{cc}
c_i = \tilde{\bldxi}^{-1} (\bldf_i) \; \forall i  & \textrm{ if } \bldq \textrm{ is integral } \\
{\tt FAILURE} & \textrm{ otherwise. }\end{array}\right.
\label{eq:LP4_output} 
\end{equation} 
\end{minipage}}
\vspace{1pc}

{\bf LP4} is capable of joint equalization and decoding, and has strong links (via Theorems \ref{prop:LP_equivalence_2} and \ref{thm:PCW_equivalence}) to the corresponding ``turbo equalizer'' based on application of the sum-product algorithm to the same factorization of the global function. Assuming for simplicity that the parity-check matrix of the LDPC code has a constant number $w_r$ of nonzero elements per row, {\bf LP4} consists of $n |\rrr|^{L+1} + m |\rrr|^{w_r-1}$ variables and $m + n|\rrr|^{L} + m w_r (|\rrr| - 1)$ constraints. Finally, note that in the case where the receiver output is integral, the LP variables $\{ w_{j,\bldbb} \}$ and $\{ q_{i,\blddd} \}$ serve as indicator functions for the local codewords and the trellis edges respectively.

\subsection{Low-complexity linear-programming receiver for the case of binary coding and modulation}
\label{low_complexity_LP_for_binary}
Note that for the case of binary coding ($\rrr = \mathbb{F}_2$) and binary modulation, a lower-complexity LP may be developed. This LP is based on the use of the ``parity polytope" of \cite{Jeroslow} which was applied to the case of linear-programming decoding of binary linear codes in \cite{Feldman}. The new LP is defined as follows, where we omit the variables $\{ w_{j,\bldbb} \}$ while introducing new variables $f_i$ for each $i \in \cU$.

\vspace{1pc}
\hspace{-3mm}\fbox{
\begin{minipage}{0.465\textwidth}
{\bf LP5: Low Descriptive Complexity Joint Equalization and Decoding (Binary Coding and Modulation)}

\vspace{1mm}

{\bf Cost Function:} 
\begin{equation}
\sum_{i \in \cU} \sum_{\blddd \in\cD^{-}} \tilde{\lambda}_i^{(\blddd)} q_{i,\blddd} 
\label{eq:cost_fn_turbo_eq_2}
\end{equation}
{\bf Constraints (Polytope $\tilde{\cQ}$):} These are given by (\ref{eq:sum_q_to_next_q}) and (\ref{eq:sum_q_to_gtilde}), together with
\begin{equation}
\forall i \in \cU, \forall \bldd \in \cD,  q_{i,\blddd} \ge 0 \; ; \quad \forall i \in \cU, \sum_{\blddd \in \cD} q_{i,\blddd} = 1 \; ,
\end{equation} 
\begin{equation}
\forall i \in \cU, \quad  0 \le f_i \le 1 \; , 
\end{equation} 
\begin{equation}
\forall j \in \cV, \forall \cF \subset \cU_j, |\cF| \mbox{ odd}, \: \sum_{i \in \cF} f_i - \sum_{i \in \cU_j \backslash \cF} f_i \le |\cF| - 1 \; ,
\end{equation} 
and
\begin{equation}
\forall i \in \cU, \quad \sum_{\blddd \in \cD: \, \ip(\blddd)=1} q_{i,\blddd} = f_i \; .
\end{equation}
{\bf Receiver Output:}
\begin{equation}
\left\{ \begin{array}{cc}
\bldc_\mathrm{out} = \bldf & \textrm{ if } \bldf \textrm{ is integral } \\
{\tt FAILURE} & \textrm{ otherwise. }\end{array}\right.
\label{eq:LP5_output} 
\end{equation} 
\end{minipage}}
\vspace{1pc}

Assuming for simplicity that the LDPC code has a constant number $w_r$ of nonzero elements per row, {\bf LP5} consists of $n(2^{L+1} + 1)$ variables and $n (2^L + 1) + m 2^{w_r-1} $ constraints. The performance of {\bf LP5} is identical to that of {\bf LP4}; this follows as a straightforward consequence of \cite[Theorem 4]{Feldman}. Note that the formulation of {\bf LP5} is equivalent to the LP problem of \cite[Theorem 13]{Kim_Pfister2} based on \cite[Definitions 8,11]{Kim_Pfister2}.

\section{Simulation Study: Joint Decoding and Equalization of Binary Coded Transmissions over an Intersymbol Interference Channel} 
\label{sec:Hamming_Proakis}

In this section we provide a simulation study of the linear-programming receiver of Section \ref{sec:equalization_decoding}. First we consider use of the binary $[7,4]$ Hamming code with BPSK modulation (constellation $\mathcal{T} = \{ -1, +1 \}$) over the Proakis B channel \cite[Chapter 10]{Proakis}; this is an intersymbol interference (ISI) channel with $L=2$ and $(h_1, \; h_2, \; h_3) = (1/\sqrt{6}, \; 2/\sqrt{6}, \; 1/\sqrt{6})$ -- the channel is static and is normalized to unity power gain. For the binary $[7,4]$ Hamming code, we use the $7 \times 7$ circulant parity-check matrix with first row $(1 \; 1 \; 1 \; 0 \; 1 \; 0 \; 0)$. The minimum AWGN pseudoweight of the Hamming code with respect to this matrix is equal to $3$, the code's minimum distance; this may be deduced by using the eigenvalue-based AWGN pseudoweight lower bound of \cite[Theorem 1]{Vontobel_Koetter_lower_bounds}. 

Due to the Proakis B channel however, the metric of importance in this context is not the minimum AWGN pseudoweight of the code, but the minimum pseudodistance of the system as defined by (\ref{eq:pseudodistance_real}). The minimum pseudodistance of the system is $d_{\mathrm{eff}}^{\mathrm{min}} = d_{\mathrm{eff}}^{(1)} = 4/3$, and the second smallest pseudodistance is $d_{\mathrm{eff}}^{(2)} = \sqrt{2}$. A complete characterization of the corresponding error events is as follows. The pseudoconfiguration corresponding to the codeword $\bldc_1 = (1 \; 1 \; 1 \; 0 \; 1 \; 0 \; 0)$ is at pseudodistance $d_{\mathrm{eff}}^{(1)} = 4/3$ from the pseudoconfiguration $\bldkappa_1$ corresponding to\footnote{Note that we must have $q_{1,\blddd} = 0$ for $\bldd \notin \left\{ 000, 100 \right\}$ and $q_{2,\blddd} = 0$ for $\bldd \notin \left\{ 000, 010, 100, 110 \right\}$, since the LP is constrained to recognize that the initial state of the channel is $\blds_0 = 00$.} 
$q_{1,100} = q_{2,110} = 1$ and
$q_{3,011} = q_{3,111} = q_{4,011} = q_{4,101} = q_{5,010} = q_{5,101} = q_{6,010} = q_{6,101} = q_{7,010} = q_{7,101} = 1/2$, for which 
\[
\bldt = (\frac{2}{\sqrt{6}}, \; -\frac{2}{\sqrt{6}}, \; -\frac{4}{\sqrt{6}}, \; -\frac{2}{\sqrt{6}}, \; 0, \; 0, \; \frac{2}{\sqrt{6}}) \; ,
\]
\[
\bldm = (\frac{2}{\sqrt{6}}, \; -\frac{2}{\sqrt{6}}, \; -\frac{3}{\sqrt{6}}, \; -\frac{1}{\sqrt{6}},  \; 0, \; 0, \; 0) \; ,
\]
\[
\bldv = (2/3, \; 2/3, \; 5/3, \; 1/3, \; 0, \; 0, \; 0)
\]
(the reader may verify using (\ref{eq:pseudodistance_real}) that the pseudodistance is $4/3$). The pseudoconfiguration corresponding to the codeword $\bldc_1$ is also at pseudodistance $d_{\mathrm{eff}}^{(2)} = \sqrt{2}$ from the pseudoconfiguration corresponding to the codeword $\bldc_2 = (1 \; 1 \; 0 \; 1 \; 0 \; 0 \; 1)$ (in this case the pseudodistance is equal to the Euclidean distance between the corresponding modulated signals). Also, the pseudoconfiguration corresponding to the codeword $\bldc_2$ is at pseudodistance $d_{\mathrm{eff}}^{(2)} = \sqrt{2}$ from the pseudoconfiguration $\bldkappa_2$ corresponding to 
$q_{1,100} = 1$, $q_{2,010} = q_{3,101} = q_{4,010} = q_{5,101} = q_{6,010} = q_{7,001} = 2/3$ and $q_{2,110} = q_{3,011} = q_{4,101} = q_{5,010} = q_{6,001} = q_{7,100} = 1/3$, for which 
\[
\bldt = (\frac{2}{\sqrt{6}}, \; -\frac{2}{\sqrt{6}}, \; -\frac{2}{\sqrt{6}}, \; 0, \; 0, \; \frac{2}{\sqrt{6}}, \; \frac{2}{\sqrt{6}}) \; ,
\]
\[
\bldm = (\frac{2}{\sqrt{6}}, \; -\frac{2}{3\sqrt{6}}, \; -\frac{2}{3\sqrt{6}}, \; 0,  \; 0, \; \frac{2}{3\sqrt{6}}, \; \frac{2}{\sqrt{6}}) \; ,
\]
\[
\bldv = (2/3, \; 2/9, \; 2/9, \; 0, \; 0, \; 2/9, \; 2/3) \; .
\]
\begin{figure}
\begin{center}\includegraphics[%
  width=\columnwidth, keepaspectratio]{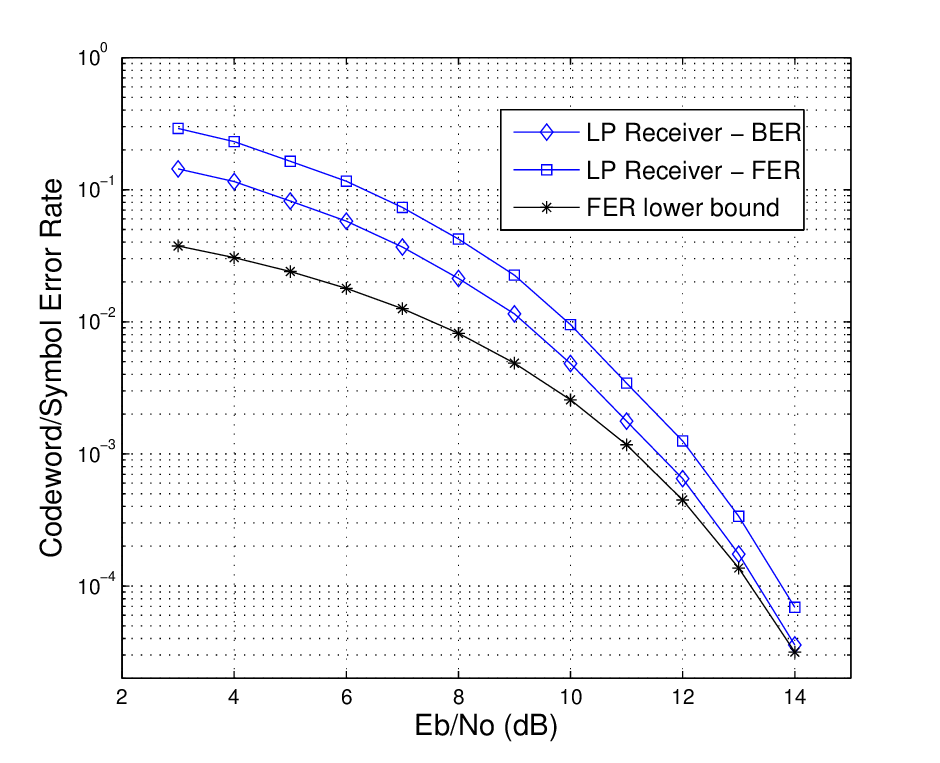}\end{center}
\caption{Bit error rate (BER) and frame error rate (FER) results for the linear-programming receiver which performs joint equalization and decoding. The plots are shown for the $[7,4]$ Hamming code and BPSK modulation over the Proakis B channel. Also plotted is the lower bound on the FER given by (\ref{eq:FER_lower_bound}). \label{cap:BER_FER_Hamming}}
\end{figure}
Similarly, the pseudoconfiguration corresponding to the codeword $\bldc_3 = (0 \; 0 \; 0 \; 1 \; 0 \; 1 \; 1)$ is at pseudodistance $d_{\mathrm{eff}}^{(1)} = 4/3$ from the pseudoconfiguration $\bldkappa_3$ corresponding to 
$q_{1,000} = q_{2,000} = 1$ and $q_{3,000} = q_{3,100} = q_{4,010} = q_{4,100} = q_{5,010} = q_{5,101} = q_{6,010} = q_{6,101} = q_{7,010} = q_{7,101} = 1/2$, for which 
\[
\bldt = (\frac{4}{\sqrt{6}}, \; \frac{4}{\sqrt{6}}, \; \frac{4}{\sqrt{6}}, \; \frac{2}{\sqrt{6}}, \; 0, \; 0, \; -\frac{2}{\sqrt{6}}) \; ;
\]
\[
\bldm = (\frac{4}{\sqrt{6}}, \; \frac{4}{\sqrt{6}}, \; \frac{3}{\sqrt{6}}, \; \frac{1}{\sqrt{6}},  \; 0, \; 0, \; 0) \; ;
\]
\[
\bldv = (8/3, \; 8/3, \; 5/3, \; 1/3, \; 0, \; 0, \; 0) \; .
\]
The pseudoconfiguration corresponding to the codeword $\bldc_3$ is also at pseudodistance $d_{\mathrm{eff}}^{(2)} = \sqrt{2}$ from the pseudoconfiguration corresponding to the codeword $\bldc_4 = (0 \; 0 \; 1 \; 0 \; 1 \; 1 \; 0)$, which in turn is at pseudodistance $d_{\mathrm{eff}}^{(2)} = \sqrt{2}$ from the pseudoconfiguration $\bldkappa_4$ corresponding to 
$q_{1,000} = 1$, $q_{2,100} = q_{3,010} = q_{4,101} = q_{5,010} = q_{6,101} = q_{7,110} = 2/3$ and $q_{2,000} = q_{3,100} = q_{4,010} = q_{5,101} = q_{6,110} = q_{7,011} = 1/3$, for which 
\[
\bldt = (\frac{4}{\sqrt{6}}, \; \frac{4}{\sqrt{6}}, \; \frac{2}{\sqrt{6}}, \; 0, \; 0, \; -\frac{2}{\sqrt{6}}, \; -\frac{2}{\sqrt{6}}) \; ,
\]
\[
\bldm = (\frac{4}{\sqrt{6}}, \; \frac{8}{3\sqrt{6}}, \; \frac{2}{3\sqrt{6}}, \; 0,  \; 0, \; -\frac{2}{3\sqrt{6}}, \; -\frac{2}{\sqrt{6}}) \; ,
\]
\[
\bldv = (8/3, \; 4/3, \; 2/9, \; 0, \; 0, \; 2/9, \; 2/3) \; .
\]
Using this analysis, we may lower bound the frame error rate (FER) by
\begin{equation}
\label{eq:FER_lower_bound}
\mbox{FER} \ge \frac{1}{8} Q \left( \frac{d^{(1)}_{\mathrm{eff}}}{2 \sigma} \right) + \frac{1}{8} Q \left( \frac{d^{(2)}_{\mathrm{eff}}}{2 \sigma} \right) \; .
\end{equation}
Here the first term is due to the codewords $\bldc_1$ and $\bldc_3$, and the second term is due to the codewords $\bldc_2$ and $\bldc_4$. The bit error rate (BER) and frame error rate (FER) performance of the linear-programming receiver is shown in Figure \ref{cap:BER_FER_Hamming}, along with the lower bound on the FER given by (\ref{eq:FER_lower_bound}). Here the MATLAB function {\tt linprog} is used to solve the LP, and $500$ reception errors were simulated for each value of $E_b/N_0$.

\begin{figure}
\begin{center}\includegraphics[%
  width=1.0\columnwidth, keepaspectratio]{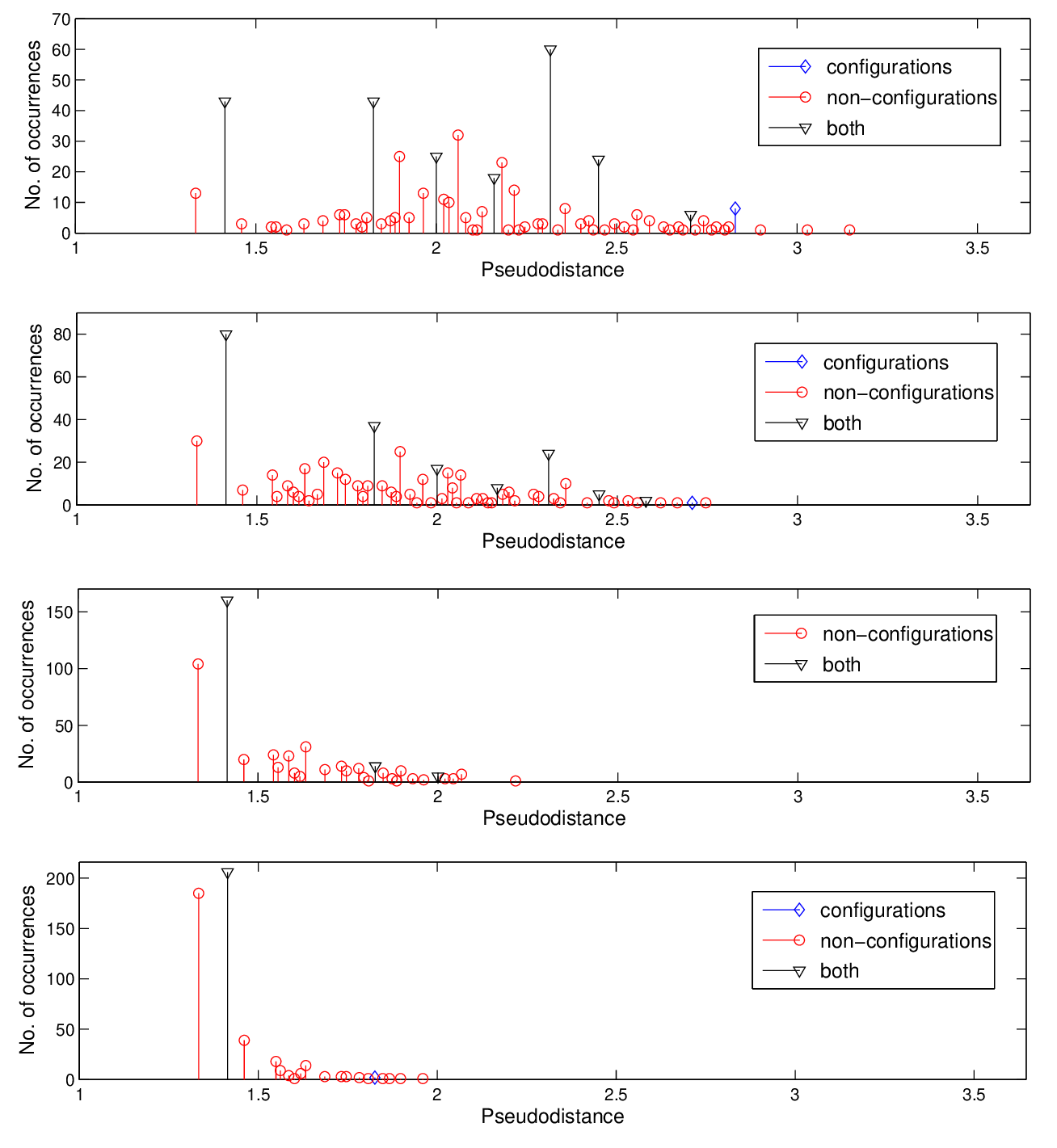}\end{center}
\caption{Measured spectra of error event pseudodistances for (top to bottom) $E_b/N_0 = 3$ dB, $7$ dB, $11$ dB and $14$ dB. The plots are for the case of $[7,4]$ Hamming coded transmission over the Proakis B channel. Each spectrum also indicates, for each value of pseudodistance $d$, whether pseudoconfigurations at distance $d$ from transmitter configurations consist of configurations only, non-configurations only, or both. \label{cap:psd_all}}
\end{figure}

The measured pseudodistance ``spectra'' are shown in Figure \ref{cap:psd_all} for four values of signal to noise ratio (SNR) spanning the simulation range; this provides a statistical record of the error events experienced by the receiver at simulated values of $E_b/N_0$ equal to $3$ dB, $7$ dB, $11$ dB and $14$ dB. Each spectrum also indicates, for each value of pseudodistance $d$, whether pseudoconfigurations at distance $d$ from transmitter configurations consist of configurations only, non-configurations only, or both. 
Note that for most transmitter configuration pairs whose pseudodistance is $d$, there also exist (configuration, non-configuration) pairs with pseudodistance $d$. It may be seen that as the SNR increases, error events at the smallest and second smallest pseudodistances, $d_{\mathrm{eff}}^{(1)} = 4/3$ (due to two error events involving non-configurations) and $d_{\mathrm{eff}}^{(2)} = \sqrt{2}$ (due to two error events involving configurations, and two involving non-configurations), begin to somewhat dominate the pseudodistance spectrum; these are the error events discussed previously in this section involving codewords $\bldc_1$, $\bldc_2$, $\bldc_3$ and $\bldc_4$. This predicts that for high SNR, the bound of (\ref{eq:FER_lower_bound}) becomes reasonably tight; tighter bounds may be developed by taking into account the nearest-neighbour pseudoconfigurations (in the pseudodistance sense) of codewords other than $\bldc_1$, $\bldc_2$, $\bldc_3$ and $\bldc_4$.

The complexity of LP decoding precludes the testing of long LDPC codes; however, as a more practical example we test a low-density code of length $n = 105$ and rate $4/7$, over two channels: {\bf CH1} refers to the Proakis B channel and {\bf CH2} refers to the length-$3$ power-normalized ISI channel given by $(h_1, \; h_2, \; h_3) = (1/\sqrt{3}, \; 1/\sqrt{3}, \; -1/\sqrt{3})$. The parity-check matrix consists of $m = 45$ rows and is equal to the right-circulant matrix
\[
\cH_{j,i} = \left\{ \begin{array}{cc}
1 & \textrm{ if } i - j \in \{ 0, 13, 48, 60 \} \\
0 & \textrm{ otherwise. }\end{array}\right. \;
\]

\begin{figure}[t]
\begin{center}\includegraphics[%
  width=1.0\columnwidth, keepaspectratio]{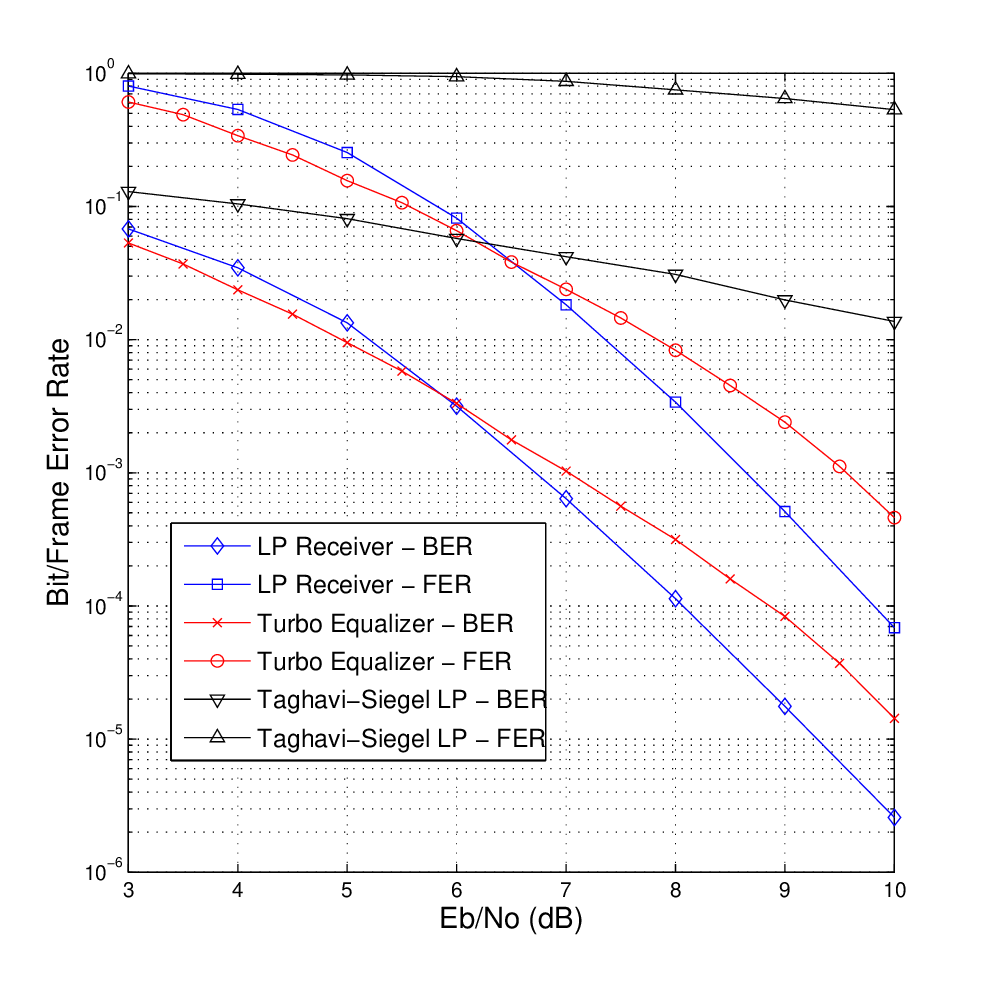}\end{center}
\caption{Bit error rate (BER) and frame error rate (FER) results for the linear-programming receiver which performs joint equalization and decoding. The $[60,105]$ low-density code is used with BPSK modulation over channel {\bf CH1}. Results are also shown for the LP receiver of Taghavi and Siegel [19], as well as for turbo equalization. \label{cap:BER_FER_1}}
\end{figure}

\begin{figure}[t]
\begin{center}\includegraphics[%
  width=1.0\columnwidth, keepaspectratio]{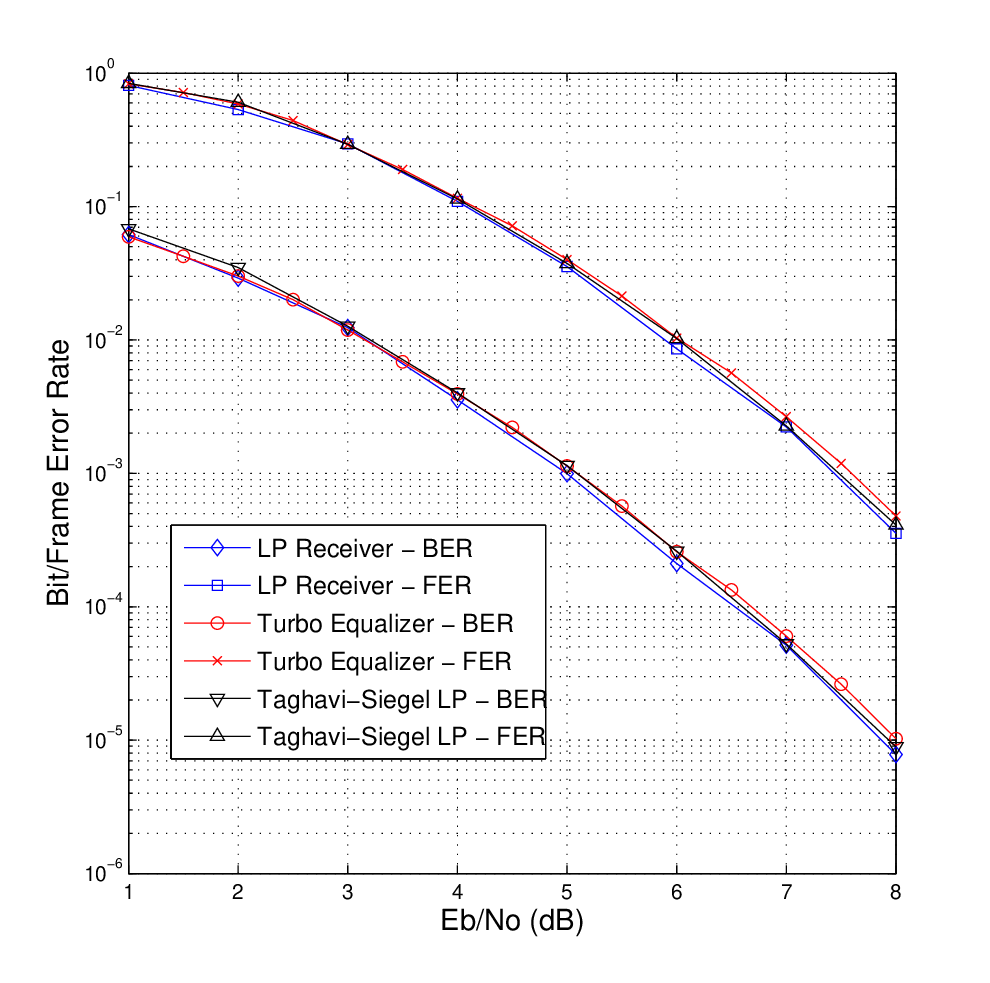}\end{center}
\caption{Bit error rate (BER) and frame error rate (FER) results for the linear-programming receiver which performs joint equalization and decoding. The $[60,105]$ low-density code is used with BPSK modulation over channel {\bf CH2}. Results are also shown for the LP receiver of Taghavi and Siegel [19], as well as for turbo equalization. \label{cap:BER_FER_2}}
\end{figure}

The results are shown in Figures \ref{cap:BER_FER_1} and \ref{cap:BER_FER_2}; also included are simulation results for two alternative receivers which perform joint equalization and decoding. The first is the classical ``turbo-equalizer'' based on the sum-product algorithm (this operates via message-passing in the factor graph of Figure \ref{cap:factor_graph}, and performs a maximum of $50$ iterations). The second is the LP-based receiver of Taghavi and Siegel presented in \cite{Siegel}. In order to achieve a practical comparison for the proposed LP receiver as well as that of \cite{Siegel}, indicator variables for the information bits were rounded to the nearest integer to form bit estimates for the BER calculation; this method was found to give much better BER performance than declaring receiver {\tt FAILURE} events, especially at low SNR.

On {\bf CH2}, all three receivers exhibit remarkably similar BER and FER performance (note that with the proposed LP receiver, the low-density coded system provides an FER gain of approximately $4$ dB over the Hamming coded system at an FER of $10^{-4}$). On {\bf CH1} the proposed LP receiver shows the best performance, outperforming even the turbo equalizer. The LP receiver of \cite{Siegel} exhibits extremely poor performance over the Proakis B channel; the reason for this that this ISI channel is not \emph{proper}. A proper channel is defined in \cite{Siegel} to be an ISI channel for which, if the system factor graph of \cite[Figure 2]{Siegel} is adopted, all cycles in the subgraph induced by the channel nodes contain an even number of ``negative'' check nodes (i.e., check nodes which have negative coefficients in the cost function, see \cite[Definitions 1 and 2, Theorem 1]{Siegel}). For {\bf CH1} however, all such check nodes are negative and the channel subgraph contains many cycles involving an odd number of negative check nodes. It may however be easily verified that channel {\bf CH1} is proper; correspondingly, the performance of the LP receiver of \cite{Siegel} is very good on {\bf CH2} and is in fact indistinguishable from that of the proposed LP receiver. As a complexity comparison, note that the LP receiver of \cite{Siegel} consists of $nL = 315$ variables and $m 2^{d-1} + 4(L-1)n = 1200$ constraints, whereas the proposed LP consists of $n(2^{L+1} + 1) = 945$ variables and $n (2^L + 1) + 8 m = 885$ constraints\footnote{Both enumerations omit complexity due to upper and lower bounds on the LP variables.}. Although a direct complexity comparison is difficult for such problems, we conclude that the complexities of both LP receivers are comparable in this case, although for longer channels the LP of \cite{Siegel} would generally be preferred in cases where the ISI channel could be proved to be proper, and the proposed LP receiver would be preferred otherwise. Note however that only a relatively small proportion of ISI channels are proper in practice.
 
The simulation estimated the minimum pseudodistance to be $d_{\mathrm{eff}}^{\mathrm{min}} = 2.3094$ for channel {\bf CH1} and $d_{\mathrm{eff}}^{\mathrm{min}} = 2.8284$ for channel {\bf CH2}; this is in accordance with the improved performance exhibited by the LP receiver on channel {\bf CH2}. Finally we remark that at present, the easiest method for computing pseudoconfigurations with low pseudodistance is simply to run the receiver at an intermediate value of SNR and record the LP outputs. However, in order to estimate the bounds of \eqref{eq:QQQ}, pseudoconfigurations with low pseudodistance would in principle need to be evaluated for every transmission, which is not feasible in practice; further research is required to address this important issue.

\section*{Conclusions and Future Work}

A general linear-programming based communication receiver design technique has been presented. It was shown that the performance of such a receiver can be characterized sharply via the concepts of \emph{maximum likelihood certificate} property, \emph{pseudoconfigurations} and \emph{system pseudodistance}. It is hoped that the results of this paper motivate further investigation into the use of system pseudodistance as a design tool for SPA as well as LP receivers. A useful tool for such LP receivers would be an efficient means of characterizing the average system pseudodistance spectrum (this is also given some attention in \cite{Kim_Pfister2}) -- such pseudodistance characterization would allow efficient system pseudodistance spectrum evaluation at low SNR to be able to provide accurate performance prediction at high SNR. However, because the pseudodistance spectrum varies depending on the transmit information word, such evaluation represents a nontrivial research problem. Another important future investigation is the application of efficient LP solvers such as the interior point methods of \cite{Wadayama_interior_pt_decoding1, Wadayama_interior_pt_decoding2, Vontobel_interior_pt_decoding} as well as the soft dual coordinate-ascent and sub-gradient based methods of \cite{Vontobel_low_complexity_LP} which may be brought to bear on this problem in order to reduce the complexity of LP receivers.

\section*{Acknowledgment}
The author would like to thank the Editor and the anonymous reviewers for their insightful technical comments which helped to improve the quality and presentation of this paper. He would also like to thank N. Boston and P. Vontobel for helpful discussions.





\begin{thebibliography} {99}

\bibitem{Berrou}
    {C. Berrou, A. Glavieux, and P. Thitimajshima,}
    {``Near Shannon limit error correcting coding and decoding: turbo-codes,''}
    {\em Proc. IEEE International Conference on Communications (ICC),} Geneva, Switzerland, pp.~1064--1070, May 1993. 

\bibitem{Gallager}
    {R. G. Gallager,}
    {``Low-density parity-check codes,''}
    {\em IRE Transactions on Information Theory,} vol. IT-8, pp.~21--28, Jan. 1962.    

\bibitem{Wiberg}
                {N. Wiberg,}
    {\em Codes and Decoding on General Graphs.} 
    Ph.D. Thesis, Link\"oping University, Sweden, 1996. 

\bibitem{Aji}
    {S. M. Aji and R. J. McEliece,}
    {``The generalized distributive law,''}
    {\em IEEE Transactions on Information Theory,} vol. 46, no. 2, pp.~325--343, March 2000.    

\bibitem{Kschischang}
    {F. R. Kschischang, B. J. Frey, and H.-A. Loeliger,}
    {``Factor graphs and the sum-product algorithm,''}
    {\em IEEE Transactions on Information Theory,} vol. 47, no. 2, pp. 498--519, Feb. 2001.    

\bibitem{Unified_design}
    {A. P. Worthen and W. E. Stark,}
    {``Unified design of iterative receivers using factor graphs,''}
    {\em IEEE Transactions on Information Theory,} vol. 47, no. 2, pp. 843--849, Feb. 2001.    

\bibitem{turbo_eq}
    {C. Douillard, M. J\'{e}z\'{e}quel, C. Berrou, A. Picart, P. Didier and A. Glavieux,}
    {``Iterative correction of intersymbol interference: turbo-equalization,''}
    {\em European Transactions on Telecommunications,} vol. 6, pp. 507--511, Sept.--Oct. 1995.  

\bibitem{Li_Poor}
    {H. Li, and H. V. Poor,}
    {``Reduced complexity joint iterative equalization and multiuser detection in dispersive {DS-CDMA} channels,''}
    {\em IEEE Transactions on Wireless Communications,} vol. 4, no. 3, pp. 1234--1243, May 2005.    

\bibitem{Goertz}
    {N. G\"{o}rtz,}
    {``On the iterative approximation of optimal joint source-channel decoding,''}
    {\em IEEE Journal on Selected Areas in Communications,} vol. 19, no. 9, pp. 1662--1670, Sept. 2001.  

\bibitem{Feldman-thesis}
    {J. Feldman,}
    {\em Decoding Error-Correcting Codes via Linear Programming.}
    Ph.D. Thesis, Massachusetts Institute of Technology, Sep. 2003.    

\bibitem{Feldman}
    {J. Feldman, M. J. Wainwright, and D. R. Karger,}
    {``Using linear programming to decode binary linear codes,''}
    {\em IEEE Transactions on Information Theory,} vol. 51, no. 3, pp.~954--972, March 2005. 

\bibitem{FSBG_SCC}
    {M. F. Flanagan, V. Skachek, E. Byrne, and M. Greferath,}
    {``Linear-programming decoding of nonbinary linear codes,''}
    {\em Proc. 7th International Conference on Source and Channel Coding (SCC 2008),} Ulm, Germany, Jan. 2008.
    Arxiv report arXiv:cs.IT/0707.4360v2, October 2007.   

\bibitem{FSBG_journal}
    {M. F. Flanagan, V. Skachek, E. Byrne, and M. Greferath,}
    {``Linear-programming decoding of nonbinary linear codes,''}
    {\em IEEE Transactions on Information Theory,} vol. 55, no. 9, pp.~4134--4154, Sep. 2009.

\bibitem{FKKR}
  {G. D. Forney, R. Koetter, F. R. Kschischang, and A. Reznik,}
  {``On the effective weights of pseudocodewords for codes defined on graphs with cycles,''}
  vol. 123 of {\em Codes, Systems, and Graphical Models,} IMA Vol. Math. Appl., ch. 5, pp.~101-112,
  Springer, 2001. 

\bibitem{KV-characterization}
  {R. Koetter, W.-C. W. Li, P. O. Vontobel, and J. L. Walker,}
  {``Characterizations of pseudo-codewords of LDPC codes,''}
  {\em Advances in Mathematics}, vol. 213, no. 1, pp. 205--229, August 2007.

\bibitem{KV-IEEE-IT}
                {P. Vontobel and R. Koetter,}
    {``Graph-cover decoding and finite-length analysis of message-passing iterative decoding of LDPC codes,''}
    to appear in {\em IEEE Transactions on Information Theory,}
    Arxiv report arXiv:cs.IT/0512078, Dec. 2005.   

\bibitem{Feldman_turbo_IRA}
    {J. Feldman and D. R. Karger,}
    {``Decoding turbo-like codes via linear programming,''}
    {\em Proc. 43-rd Annual IEEE Symposium on Foundations of Computer Science (FOCS '02),} pp. 251--260, Nov. 2002.

\bibitem{Vontobel_min_sum_LP}
                {P. Vontobel and R. Koetter,}
  {``On the relationship between linear programming decoding and min-sum decoding,''}
  {\em Proc. IEEE International Symposium on Information Theory and its Applications,} Parma, Italy, Oct. 2004.

\bibitem{Siegel}
    {M. H. Taghavi and P. H. Siegel,}
    {``Graph-based decoding in the presence of ISI,''}
    {\em IEEE Transactions on Information Theory,} vol. 57, no. 4, pp.~2188--2202, April 2011.     
	
\bibitem{Cohen}
    {A. Cohen, F. Alajaji, N. Kashyap and G. Takahara,}
    {``LP decoding for joint source-channel codes and for the non-ergodic {P}olya channel,''}
    {\em IEEE Communications Letters,} vol. 12, no. 9, pp.~678--680, September 2008.  
		
\bibitem{Kelley-Sridhara-ISIT-2006}
  {C. A. Kelley, D. Sridhara, and J. Rosenthal,}
  {``Pseudocodeword weights for non-binary LDPC codes,''}
  {\em Proc. IEEE International Symposium on Information Theory (ISIT)}, Seattle, USA, pp.~1379-1383, July 2006. 

\bibitem{Flanagan_cw_ind}
    {M. F. Flanagan,} 
    {``Codeword-independent performance of nonbinary linear codes under linear-programming and sum-product decoding,''}
    {\em Proc. IEEE International Symposium on Information Theory,} Toronto, Canada, 6--11 July 2008.     

%
\bibitem{Fossorier_on_the_equivalence}
    {M. P. C. Fossorier, F. Burkert, S. Lin and J. Hagenauer,}
    {``On the equivalence between {SOVA} and {M}ax-{L}og-{MAP} decodings,''}
    {\em IEEE Communications Letters,} vol. 2, no. 5, pp.~137--139, May 1998. 

\bibitem{Jeroslow}
  {R. G. Jeroslow,}
  {``On defining sets of vertices of the hypercube by linear inequalities,''}
  {\em Discrete Mathematics}, vol.~11, no.~2, pp.~119--124, 1975.

\bibitem{Proakis}
    {J. G. Proakis,} 
    {\em Digital Communications.}
    Fourth edition, McGraw-Hill, 2001. 

\bibitem{Vontobel_Koetter_lower_bounds}
  {P. O. Vontobel and R. Koetter,}
  {``Lower bounds on the minimum pseudo-weight of linear codes,''}
  {\em Proc. IEEE International Symposium on Information Theory}, Chicago, USA, pp.~67, June/July 2004.

\bibitem{Kim_Pfister}
  {B.-H. Kim and H. D. Pfister,}
  {``On the joint decoding of {LDPC} codes and finite-state channels via linear programming,''}
  {\em Proc. IEEE International Symposium on Information Theory}, Austin, Texas (USA), June 13--18, 2010. 

\bibitem{Kim_Pfister2}
  {B.-H. Kim and H. D. Pfister,}
  {``Joint decoding of {LDPC} codes and finite-state channels via linear-programming,''}
  submitted to {\em IEEE Journal of Selected Topics in Signal Processing (Special Issue on Soft Detection for Wireless Transmission)}, Feb. 2011. 

\bibitem{Wadayama_interior_pt_decoding1}
  {T. Wadayama,}
  {``Interior point decoding for linear vector channels,''}
  {\em Proc. IEEE International Symposium on Information Theory,} Toronto, Canada, 6--11 July 2008. 

\bibitem{Wadayama_interior_pt_decoding2}
  {T. Wadayama,}
  {``An LP decoding algorithm based on primal path-following interior point method,''}
  {\em Proc. IEEE International Symposium on Information Theory}, Seoul, Korea, pp.~389--393, June 28--July 3, 2009. 

\bibitem{Vontobel_interior_pt_decoding}
  {P. O. Vontobel,}
  {``Interior-point algorithms for linear-programming decoding,''}
  {\em Proc. Information Theory and Applications Workshop,} San Diego, USA, pp.~433--437, Jan. 2008.

\bibitem{Vontobel_low_complexity_LP}
  {P. O. Vontobel and R. Koetter,}
  {``Towards low-complexity linear-programming decoding,''}
  {\em Proc. 4th International Conference on Turbo Codes and Related Topics}, Munich, Germany, Apr. 3--7, 2006.

\end{thebibliography}
\end{document}